\documentclass[fleqn,usenatbib]{mnras}

\usepackage[T1]{fontenc}

\DeclareRobustCommand{\VAN}[3]{#2}
\let\VANthebibliography\thebibliography
\def\thebibliography{\DeclareRobustCommand{\VAN}[3]{##3}\VANthebibliography}

\usepackage{graphicx}	
\usepackage{amsmath}	
\usepackage{amssymb}	
\usepackage[export]{adjustbox}

\usepackage{morefloats}
\usepackage{newtxtext,newtxmath}

\def\ha{H$\alpha$}
\def\hb{H$\beta$}

\def\sfrg{SFR$_{Gas}$ }
\def\sfrs{SFR$_\star$ }
\def\sfrgd{$\rm \Sigma SFR_{Gas}$}
\def\sfrsd{$ \rm \Sigma SFR_\star$}

\def\str{{\sc starlight}}
 
\def\st{{\sc starlight}}

\newcommand\footnoteref[1]{\protected@xdef\@thefnmark{\ref{#1}}\@footnotemark}

\title[SFR in AGN hosts]{Determining star-formation rates in Active Galactic Nuclei hosts via stellar population synthesis}
\author[Riffel, R. et al.]{Rog\'erio Riffel$^{1,2}$\thanks{E-mail: riffel@ufrgs.br},
Nicolas D. Mallmann$^{1,2}$,
Gabriele S. Ilha$^{2,3}$,
Thaisa Storchi-Bergmann$^{1,2}$,
\newauthor
Rogemar A. Riffel$^{2,3}$,
Sandro B. Rembold$^{2,3}$,
Dmitry Bizyaev$^{4,5}$,
Janaina C. do Nascimento$^{1,2}$,
\newauthor
Jaderson S. Schimoia$^{2,3}$,
 Luiz N. da Costa,$^{2,6}$
Nicholas Fraser Boardman $^{7}$,
\newauthor
M\'ed\'eric Boquien$^{8}$,
Guilherme S. Couto$^{8}$,
\\
$^{1}$ Departamento de Astronomia, Instituto de F\'\i sica, Universidade Federal do Rio Grande do Sul, CP 15051, 91501-970, Porto Alegre, RS, Brazil \\
$^{2}$Laborat\'orio Interinstitucional de e-Astronomia - LIneA, Rua Gal. Jos\'e Cristino 77, Rio de Janeiro, RJ - 20921-400, Brazil\\
$^{3}$ Departamento de F\'\i sica, Centro de Ci\^encias Naturais e Exatas, Universidade Federal de Santa Maria, 97105-900, Santa Maria, RS, Brazil \\ 
$^4$ Apache Point Observatory and New Mexico State
University, P.O. Box 59, Sunspot, NM, 88349-0059, USA\\
$^5$ Sternberg Astronomical Institute, Moscow State
University, Moscow, Russia\\
$^{6}$Observat\'orio Nacional - MCT, Rua General Jos\'e Cristino 77, Rio de Janeiro, RJ - 20921-400, Brazil\\
$^{7}$ Department of Physics \& Astronomy, University of Utah, Salt Lake City, UT, 84112, USA\\
$^{8}$ Centro de Astronom\'\i a (CITEVA), Universidad de Antofagasta, Avenida Angamos 601, Antofagasta, Chile.\\
}

\date{Accepted XXX. Received YYY; in original form ZZZ}

\pubyear{2020}

\begin{document}
\label{firstpage}
\pagerange{\pageref{firstpage}--\pageref{lastpage}}
\maketitle

\begin{abstract}
The effect of active galactic nuclei (AGN) feedback on the host galaxy, and its role in quenching or enhancing star-formation, is still uncertain due to the fact that usual star-formation rate (SFR) indicators -- emission-line luminosities based on the assumption of photoionisation by young stars -- cannot be used for active galaxies as the ionising source is the AGN. We thus investigate the use of SFR derived from the stellar population and its relation with that derived from the gas for a sample of 170 AGN hosts and a matched control sample of 291 galaxies. We compare the values of SFR densities obtained via the \ha\ emission line (\sfrgd) for regions ionised by hot stars according to diagnostic diagrams with those obtained from stellar population synthesis (\sfrsd) over the last 1 to 100~Myr. We find that the \sfrsd over the last 20~Myrs closely reproduces the \sfrgd, although a better match is obtained via the transformation:  $\mbox{log(\sfrsd)} = (0.870\pm0.004)\mbox{log(\sfrgd)} +(0.007\pm0.006)$ (or $\mbox{log(\sfrgd)} = (1.149\pm0.005)\mbox{log(\sfrsd)}  - (0.008\pm0.008)$), which is valid for both AGN hosts and non-active galaxies. We also compare the reddening obtained via the gas \ha/\hb\ ratio  with that derived via the full spectral fitting in the stellar population synthesis.  We find that the ratio between the gas and stellar extinction is in the range  2.64 $ \le A_{Vg}/A_{V\star} \le$ 2.85, in approximate agreement with previous results from the literature, obtained for smaller samples. We interpret the difference as being due to the fact that the reddening of the stars is dominated by that affecting the less obscured underlying older population, while the reddening of the gas is larger as it is associated to a younger stellar population buried deeper in the dust.

\end{abstract}




\section{Introduction}

Present day galaxies display a wide range of 
luminosities, sizes, stellar population properties, structure, kinematics, and gas content, being the endpoint of 
a $\sim$13.8 Gyr long process \citep{Aghanim+20}. 
These properties have been shaped by a series of processes, and, according to them, galaxies can roughly be divided in passive and star-forming. The passive galaxies are not actively forming stars and host a red and old stellar population, while the star-forming galaxies are blue, hosting large fractions of young stellar populations. Such bi-modal behaviour is observed even at high redshifts ($z > 2.5$) where populations of passive galaxies are observed \citep[e.g.][]{Muzzin+13,Brammer+09}. 

The bi-modality of galaxies has been verified in a number of studies over the years 
\citep[e.g.][]{Baldry+04,Wetzel+12,Kauffmann+03a,Noeske+07,Wel+14}. However, it is not yet clear which mechanisms are driving the shutting down of star formation and transforming the blue star-forming spiral galaxies into {\it red-and-dead} galaxies. A major challenge in modern astrophysics is to determine the nature of the physical mechanism quenching star formation in galaxies. 

One mechanism that has been invoked by a number of studies is the feedback of active galactic nuclei (AGN). AGN feedback can quench star formation by heating and/or (re)moving the gas. AGN outflows are often considered as negative feedback processes 
that suppress star-formation \citep[e.g.][and references therein]{Granato+04,Fabian+12,King+15,Zubovas+17a,Trussler+20}. On the other hand, some models and simulations suggest that these outflows and jets can in some cases compress the galactic gas, and therefore act as a catalyzer and boosting the star-formation \citep[e.g.][]{Rees+89,Hopkins+12,Nayakshin+12,Bieri+16,Zubovas+13,Zubovas+17a} and even form stars inside the outflow \citep[e.g.][for an observational example see \citet{Gallagher+19}, and references therein]{Ishibashi+12,Zubovas+13,El-Badry+16,Wang+18}.

Cosmological simulations \citep[e.g.][]{Springel+05,Vogelsberger+14,Crain+15} performed without the inclusion of feedback effects are not able to reproduce the galaxy luminosity function (at both the low and high-luminosity ends), and also underestimate the ages of the most massive galaxies \citep[see Figs.~8~and~10 of][]{Croton+06}. It thus seems that effective feedback is required to reproduce the galaxy properties, but simulations can only provide limited insight into the nature and source of the feedback processes \citep[e.g. AGN or SN dominated][]{Schaye+15}. This is because there are still not enough observational constraints on these processes, and in particular, in the case of AGN, in order to verify how quenched is the star formation it is necessary to robustly quantify the 
star formation rates (SFR) of the hosts in the vicinity of the AGN.  
Both AGN activity and star-formation (SF) are regulated by the amount of available gas in the host galaxy. The relation between the  gas and SF is relatively well-characterised by previous studies \citep[e.g.][]{Kennicutt+12,Barrera-Ballesteros+20,Lin+19,Zhuang+20a}. Simulations should be able to correctly reproduce the observed SF; this, however, will depend on the efficiency of feedback processes in the interstellar medium (ISM). But resolving such processes is not yet possible in simulations of cosmological volumes \citep{Schaye+15}. In addition, in current models, the feedback processes are included in an {\it ad-hoc} manner \citep{Weinberger+17,Nelson+19}, being activated by a threshold luminosity.

While there are many calibrators to determine the SFR for non-active galaxies  \citep[for a review see][]{KennicuttJr.+98,Kennicutt+12}, in the case of AGN this is a very difficult task, since the emitting gas is ionised by the AGN radiation. Therefore, SFR indicators and equations calibrated with stellar photoionization prescriptions can not be used in this context.  Efforts have been made using  far infrared luminosities as star formation indicators \citep[e.g.][and references therein]{Kennicutt+09,Rosario+13,Rosario+16,Rosario+18} or mid-infrared neon ([\ion{Ne}{ii}] 12.81$\mu$m and [\ion{Ne}{iii}] 15.56$\mu$m) emission lines \citep{Zhuang+19}. The infrared observations used in these works are obtained using large apertures, including the whole galaxy emission.  As discussed in \citet{Rosario+16} despite finding a good correlation between Far-infrared emission  and SFR (which can be attributed to leaking of the SF regions to the whole galaxy or that the SF is constant over several hundreds of Myr) a substantial component of the cold dust luminosity that is associated with a diffuse interstellar radiation field can come from evolved stars. Even if the contribution from an old age stellar population is not dominant, the far-infrared may not be a good tracer of SF since dust can be heated by stars that are older than a few hundred Myr \citep{Kennicutt+09,Hao+11,Rosario+16}.  Other recent effort on the bluer part of the spectrum has  been made using a re-calibration [\ion{O}{ii}] $\lambda$3727 emission line \citep{Zhuang+20,Zhuang+19}, which is metallicity and electron density  dependent.

From the above it is clear that it  is thus necessary to find an independent way to obtain SFRs in AGN hosts.  A powerful technique to disentangle the components summing up to a galaxy spectral energy distribution (SED) is stellar populations synthesis \citep[e.g.][and references therein]{CidFernandes+04,CidFernandes+05,Riffel+09,Riffel+08,Walcher+11,CidFernandes+18,Baldwin+18,Salim+18,Peterken+20}. The synthesis is based on the simultaneous fit of different proportions of composite or simple stellar populations (SSPs) templates from a base of such spectra to the observed spectrum.  In the case of the \str\ fitting code \citep{CidFernandes+04,CidFernandes+05,CidFernandes+18}, it returns the values of the gas mass rate that has been converted into stars throughout the galaxy life as well as this rate for each single SSP included in the base. These values can, therefore, be used to compute the SFR via stellar population synthesis, \sfrs. In fact, this was already applied in \citet{Asari+07} using single fiber Sloan Digital Sky Survey (SDSS) data and older stellar population models generations \citep[they used ][ models]{Bruzual+03a}.

From the discussion above it is clear that it is of utmost importance to characterise the SFR of AGN hosts -- in particular in the narrow-line region (NLR) and extended narrow-line region (ENLR), that are photoionised by the AGN. And in order to verify the actual role of the AGN on the SFR it is necessary to compare the SFR values obtained for the NLR and ENLR with those obtained for a matched control sample of non-active galaxies at similar distances from the nucleus, as well as with predictions from simulations. In \citet{Rembold+17} we described the method we have used to select our sample of AGN and a matched control galaxy sample from the Mapping Nearby Galaxies at APO \citep[MaNGA, ][]{Bundy+15a} survey. 
Here, we have used an updated sample of 170 AGN and 291 controls (Deconto-Machado, {\it in preparation}), selected as in \citet{Rembold+17} to compute SFR indicators from the gas -- using diagnostic diagrams to separate gas ionised by AGN and by hot stars -- and from the stellar population synthesis and investigate the relation between them. We propose equations to relate the SFR densities obtained via emission lines, \sfrgd\ with that obtained via stellar population synthesis, \sfrsd, allowing to obtain one from the other. 

This paper is structured as follows: in section~\ref{sec:data} we present the updated samples. The methods used to determine the SFR are described in section~\ref{sec:sfr}. The results are presented and discussed in section~\ref{sec:res_disc}  and conclusions are made in section~\ref{sec:conclusions}. We have used throughout the paper
$H_0=73$\,km\,s$^{-1}$\,Mpc$^{-1}$.

\section{data }\label{sec:data}

The data used here were obtained from the Mapping Nearby Galaxies at Apache Point Observatory (MaNGA) survey \citep{Bundy+15a}.  MaNGA is part of the fourth generation Sloan Digital Sky Survey (SDSS IV). The survey has provided optical spectroscopy ($3600$\,{\AA}-$10400$\,{\AA}) of $\sim 10,000$ nearby galaxies (with $\langle z\rangle\,\approx\,0.03$). The observations were carried out with fiber bundles of different sizes (19-127 fibers) covering a field of $12''$ to $32''$ in diameter. MaNGA observations are divided into ``primary'' and ``secondary'' targets, the former are observed up to $1.5$ effective radius ($R_e$) while the latter is observed up to $2.5\,R_e$. For more details, see \citet{Drory+15,Law+15,Yan+16a,Yan+16}.

The sample used in this work is an update of our previous MaNGA AGN hosts and matched non-active control galaxies \citep{Rembold+17}. The control sample was selected in order to match the AGN hosts in terms of stellar mass, redshift, visual morphology and inclination \citep[for details see][]{Rembold+17}.  After the release of the MaNGA Product Launch 8 \citep[MPL-8, ][]{Aguado+19,Blanton+17,Bundy+15a,Belfiore+19,Westfall+19,Cherinka+19,Wake+17,Law+15,Law+16,Yan+16,Yan+16a,Drory+15,Gunn+06,Smee+13}, the number of observed AGN with MaNGA has grown to 170 AGNs using the same criteria as in \citet{Rembold+17}. For each AGN, we have also selected two control galaxies. Since more than one AGN host can share the same control galaxy, this inactive sample is composed by 291 sources. Both AGN and  control samples are located in the redshift range $0.02\lesssim z \lesssim 0.15$, and their typical stellar masses
are of the order $10^{10.5}-10^{11}M_\odot$. Most AGN in our sample (64\%) are low-luminosity, presenting [\ion{O}{iii}]\,$\lambda$5007\AA\   luminosities below $3.8\times 10^{40}\,\mbox{erg}\,\mbox{s}^{-1}$.

Analysis of the morphological classification of galaxies in our sample with the Galaxy Zoo database \citep{Lintott+08,Lintott+11} reveals that our updated AGN sample contains 57 early-type (33.5 percent), 87 late-type (51.2 percent), 3 merger galaxies (1.8 percent), and  23 objects (13.5 percent) without classification. The control sample is composed of 125 early-type (36.8 percent), 182 late-type (53.5 percent), 4 merger galaxies (1.2 percent), and 29 objects (8.5 percent) whose classifications are uncertain. Regarding nuclear activity, 63.6 percent of the AGN host sample presents Seyfert nuclei, while the other 36.4 percent are 
Low-ionization nuclear emission-line region (LINER) sources. For more  details on the updated sample properties see Deconto-Machado ({\it in preparation}).

 For all galaxies in our sample, gas and stellar population parameters have been derived from the MaNGA IFU optical spectra datacubes. We refer to \citet{Bundy+15a} for details on the MaNGA spectroscopic data, like spectral resolution, spatial coverage and pixel scale.

\section{star-formation rates } \label{sec:sfr}

We have obtained the star-formation rate surface densities for the gas \sfrgd\ and for the stars \sfrsd\ for each spaxel of the datacubes dividing the corresponding SFR values in units of solar masses per year ($M_\odot$\,yr$^{-1}$) by the area of each spaxel in kpc$^2$.

Our goal is to compare SFR values obtained from the gas emission lines to those obtained from stellar population synthesis. As the prescriptions for the gas are based on the assumption that it is photoionised by young, hot stars, this comparison needs to be done only for spaxels in which the gas is indeed ionised by stars. We have thus used optical diagnostic diagrams (see \S~\ref{validation}) to isolate the spaxels whose emission is produced by photoionisation by the radiation of young stars.

We also need to be sure that the signal-to-noise ratio of the data from each spaxel is high enough to allow reliable measurements. In summary, in order to ensure that the measurements are accurate, we have subjected the results obtained for each spaxel to the validation criteria listed below.

\subsection{Spaxels validation}
\label{validation}

We considered results from a spaxel to be valid only if they match the following criteria:
\begin{itemize}
    \item The mean value of the signal-to-noise ratio (SNR) in the continuum window between 5650\AA\ and 5750\AA\ is $\geq$ 10. This was applied in order to ensure that the stellar population fits are reliable \citep{CidFernandes+04,CidFernandes+05,Riffel+09};
    \item The \ha\ Equivalent width (EW) is larger than 10~\AA\ and the \hb\ EW is larger than 3~\AA. This is necessary in order to avoid spaxels that could have large contribution from other ionising sources such as post-AGB stars;
    \item The following relation between line ratios is obeyed: $\log([\ion{O}{iii}]\lambda 5007/$\hb$) < 0.61/(\log([\ion{N}{ii}]\lambda 6583/$\ha${}) - 0.05) + 1.3$ \citep{Kauffmann+03b}. This is based on diagnostic diagrams to ensure that only star-forming emission spaxels are being used in the calculation of the SFR.
\end{itemize}

By using these selection criteria we make sure that we are only  using results obtained for spaxels that have a good stellar population fit and the gas is photoionised by hot young stars.

\subsection{SFR from stellar population synthesis}

Our first step was to perform a full spectral fitting stellar population synthesis on our datacubes. We used the \st\ fitting code \citep{CidFernandes+05,CidFernandes+18} which combines the spectra of a base of $N_{\star}$ simple stellar population (SSP) template spectra $b_{j,\lambda}$, in different proportions, in order to reproduce the observed spectrum $O_\lambda$. For this comparison the modelled spectra $M_\lambda$ are normalised at an user defined wavelength ($\lambda_0$). The reddening is given by the term $r_\lambda = 10^{-0.4 (A_\lambda - A_{\lambda_0})}$, weighted by the population vector $x_j$ (which represents the fractional contribution of the $j$th SSP to the light at the normalisation wavelength $\lambda_0$), and convolved with a Gaussian distribution $G(v_\star, \sigma_\star)$ to account for velocity shifts $v_{\star}$, and velocity dispersion $\sigma_{\star}$. 

Each model spectrum can be expressed as:

\begin{equation}
	M_\lambda = M_{\lambda_0} \left[ \sum_{n=1}^{N_\star} x_j\,b_{j,\lambda}\,r_\lambda \right] \otimes G(v_\star, \sigma_\star),
\end{equation}

\noindent where $M_{\lambda_0}$ is the flux of the synthetic spectrum at the wavelength $\lambda_0$. To find the best parameters for the fit, the code searches for the minimum of $\chi^2 = \sum_{\lambda_i}^{\lambda_f} [(O_\lambda - M_\lambda) \omega_\lambda]^2$, where $\omega_\lambda$ is the inverse of the error, using a simulated annealing plus Metropolis scheme. We normalised our data at $\lambda_0$, adopted to be the mean value between 5650\AA\ and 5750\AA. The reddening law we have used was that of \citet{Cardelli+89a} and the synthesis was performed for the spectral range from $3700$\,{\AA} to $6900$\,{\AA}.

The SSPs base set we use is the $GM$ described in \citet{CidFernandes+13,CidFernandes+14} that is constructed using the {\sc Miles} \citep{Vazdekis+10} and \citet{GonzalezDelgado+05} models. We have updated it with the {\sc Miles} V11 models \citep{Vazdekis+16a}.  We used 21 ages (t= 0.001, 0.006, 0.010, 0.014, 0.020, 0.032, 0.056, 0.1, 0.2, 0.316, 0.398, 0.501 0.631, 0.708, 0.794, 0.891, 1.0, 2.0, 5.01, 8.91 and 12.6 Gyr) and four metallicities (Z= 0.19, 0.40, 1.00 and 1.66 Z$_\odot$). We have also added to the specgral base a power law of the form $F_{\nu} \propto \nu^ {-1.5}$ to account for the contribution of a possible AGN continuum (observed directly or as scattered light). 

Since \st\ is not prepared to handle with datacubes we have used our in house software {\sc megacube} \citep{Mallmann+18a}. 
This  code wraps \st\ to deal with its numerous input and output files involved with IFU data. Each spaxel requires an ASCII file and generates another one. For each galaxy, thousands of files are organised and extracted to a coherent data-cube to be subsequently analysed. The code can also be used for a multitude of functions since it was developed with modular capabilities, i.e., parts of the software can be  changed, swapped or removed depending on the scientific goals.
Besides allowing us to easily prepare and fit the stellar populations in datacubes its modular approach allows to use {\sc megacube} to generate maps for the direct and indirect \st\ fitting products. See \citet{Mallmann+18a} for further details.

One of the data products computed by {\sc megacube} is the star formation rate obtained from the stellar fit ($SFR_\star$) over an user-defined age interval ($\Delta t = t_{j_f}-t_{j_i}$). This can be computed since the SSPs model spectra are in units of L$_{\odot}$ \AA$^{-1} M_{\odot}^{-1}$, and the observed spectra ($O_{\lambda}$) are in units of $\rm erg/s/cm^2/$\AA\ (for details see the \str\ manual\footnote{\label{note1} http://www.starlight.ufsc.br/ }). The SFR$_\star$ over the chosen $\Delta t$  can be computed assuming that the mass of each base component ($j$) which has been processed into stars can be obtained as:
\begin{equation}
    M_{\star,j}^{\rm ini} = \mu_{j}^{\rm ini} \times \frac{4\pi d^2}{3.826\times 10^{33}},
\end{equation}
where $M_{\star,j}^{\rm ini}$ is given in $M_\odot$, $\mu_{j}^{\rm ini}$ represents the mass that has been converted into stars for the $j$-th element and its flux. This parameter is given in $M_\odot\,{\rm erg s^{-1} cm^{-2}}$; $d$ is the distance to the galaxy in cm and 3.826$\times10^{33}$ is the Sun's luminosity in erg\,s$^{-1}$. Thus, the SFR over the $\Delta t$ as defined above
can be obtained from the equation:
\begin{equation}
    {\rm SFR_\star} = \frac{\sum_{j_i}^{j_f} M_{\star,j}^{\rm ini}}{\Delta t}.
\end{equation}

\noindent For more details, see the \st\ manual\textsuperscript{\ref{note1}}, and for an application example see \citet{Asari+07,Riffel+20}.

\subsection{SFR from \ha\ emission-line fluxes}

A very common approach to determine SFR from the gas emission ($\rm SFR_{Gas}$)  is using hydrogen recombination emission-line fluxes \citep[see, ][ for a review]{KennicuttJr.+98,Kennicutt+12}. Therefore, we have used the {\sc megacube} absorption free emission-line datacubes to fit the emission lines. This procedure was done using the {\sc ifscube}\footnote{https://ifscube.readthedocs.io/en/latest/} tool:  a python package designed to perform analysis tasks in data cubes. This code allows to fit emission lines with Gaussian functions (among other options), allowing to constraint kinematics and line fluxes ratios in a very robust and easy way \citep{Ruschel-Dutra+20}.

We use the {\sc ifscube} Python package to fit the emission-line profiles of H$\beta$, [O\,{\sc iii}]$\lambda\lambda$4959,5007, He\,{\sc i}$\lambda$5876, [O\,{\sc i}]$\lambda$6300, H$\alpha$ [N\,{\sc ii}]$\lambda\lambda$6548,6583 and [S\,{\sc ii}]$\lambda\lambda$6716,6731. We fit the spectra after the subtraction of the underlying stellar population  contribution derived in the previous section. The line profiles are fitted with Gaussian curves by adopting the following constraints: (i) the width and centroid velocities of emission lines from the same parent ion are tied; (ii) the [O\,{\sc iii}]$\lambda$5007/$\lambda$4959 and  [N\,{\sc ii}]$\lambda$6583/$\lambda$6548 flux ratios are fixed to their theoretical values of 2.98 and 3.06, respectively; (iii) the centroid velocity is allowed to vary from --300 to 300 km\,s$^{-1}$ for [\ion{S}{ii}] lines and --350 to 350 km\,s$^{-1}$ for the other lines relative to the velocity obtained from the redshift of each galaxy (listed in the MaNGA Data Analysis Pipeline - DAP); and (iv) the observed velocity dispersion of all lines is limited to the range 40--300 km\,s$^{-1}$. In addition, we include a first order polynomial to reproduce the local continuum.

Since emission lines effectively re-emit the photons absorbed from the integrated stellar Lyman continuum, they provide a direct probe of the population of young, massive stars. Maps for the $\rm SFR_{Gas}$  were obtained with the following equation from \citet{KennicuttJr.+98}:
\begin{equation}
    \mathrm{SFR_{Gas}} (M_\odot / \mathrm{yr}) = 7.9 \times 10^{-42} L(\mathrm{H}\alpha) \mathrm{(ergs/s)},
\end{equation}
were $L(\mathrm{H}\alpha)$ is the reddening corrected \ha\ luminosity.

\subsection{Stellar population and nebular reddening}

\st\ models the reddening of the integrated stellar continuum as a foreground dust screen and it is parameterized by the extinction in the $V$-band, $A_V$ using the reddening law of \citet[CCM][]{Cardelli+89a}. 

The gas reddening was obtained considering Case B recombination at $\rm T_e$~=~10000\,K \citep{KennicuttJr.+98}, and the corresponding color excess $E(B-V)$ can be obtained as follows \citep[see][]{Calzetti+00,Dominguez+13}:
\begin{eqnarray}
    E (B-V) & = & \frac{E(\mathrm{H}{\beta} - \mathrm{H}{\alpha}) }{ f_{\lambda} ( \mathrm{H}{\beta}) - f_{\lambda} ( \mathrm{H}{\alpha})} \\
     & = &
    \frac{2.5}{R_{\lambda} \left(f_{\lambda} ( \mathrm{H}\beta) - f_{\lambda} ( \mathrm{H}\alpha)\right)} \left[\frac{(F_{\mathrm{H}\alpha}/F_{\mathrm{H}\beta})^{obs}}{(F_{\mathrm{H}\alpha}/F_{\mathrm{H}\beta})^{int}}\right],
\end{eqnarray}

\noindent where $ f_{\lambda} (\mathrm{H}{\alpha})$ and $f_{\lambda} ( \mathrm{H}{\beta})$ are the reddening curve values at the $\mathrm{H}{\alpha}$ and $ \mathrm{H}{\beta}$ wavelengths, which for the CCM's reddening law are $ f_{\lambda} (\mathrm{H}{\alpha})=0.818$ and $f_{\lambda} (\mathrm{H}_{\beta})=1.164$. 
 
 Adopting $ R_\lambda = R_V = 3.1$ and the theoretical line ratio of ${F_{H\alpha}/F_{H\beta}}=2.86$ for case B \ion{H}{i} recombination for an electron temperature of $T_e = 10\,000 $~K and electron density of $N_e = 100~\mbox{cm}^{-3}$ \citep{Osterbrock+06}, we obtain:

\begin{equation}
    A_V = 7.22~\log \left(\frac{(F_{\mathrm{H}\alpha}/F_{\mathrm{H}\beta})^{obs}}{2.86} \right).
\end{equation}

The intrinsic flux ($ F_{\rm int}^{\lambda}$) of an emission line is then related to the observed one ($F_{\rm obs}^{\lambda}$) by the following equation: 
\begin{eqnarray}
     F_{\rm int}^{\lambda} & = & F_{\rm obs}^{\lambda}10^{0.4A_{\lambda}} \\
    \noindent  & = &  F_{\rm obs}^{\lambda}10^{0.4R_{\lambda} E(B-V)},
\end{eqnarray}
\noindent where $A_{\lambda}$ is the extinction at wavelength $\lambda$ and $R_{\lambda}$ is the extinction curve index from  \citet{Cardelli+89a}. The equation above was used to correct the H$\alpha$ emission-line flux used in the calculation of the H$\alpha$ luminosity and \sfrg.

\section{Results and Discussion }\label{sec:res_disc}

\subsection{Comparison between $SFR_{Gas}$ and $SFR_*$} 

In order to compare the SFR values obtained for all galaxies that are at different distances, we have calculated the SFR surface densities \sfrsd\ and \sfrgd\ by dividing the obtained SFRs values for each spaxel by its area in $kpc^{2}$. The resulting values are in units of M$_\odot$ $\rm yr^{-1}$ $\rm kpc^{-2}$.

\begin{figure*}
    \centering
     \includegraphics[width=0.95\textwidth,height=0.95\textheight, keepaspectratio]{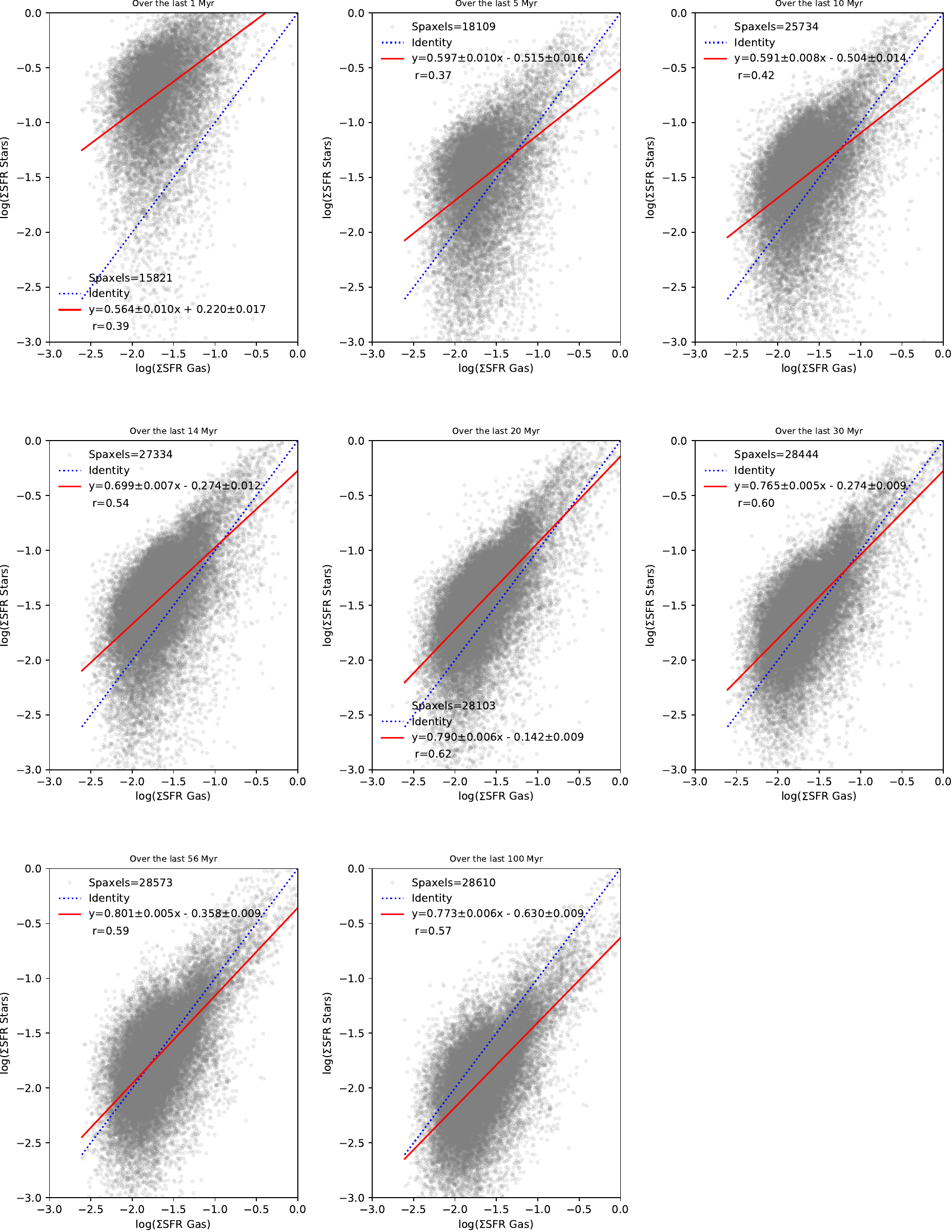}
    \caption{Comparison of \sfrgd with \sfrsd in logarithm units of M$\odot$\,yr$^{-1}$\,kpc$^{-2}$ over the last  1,  5, 10, 14, 20, 30, 56 and 100~Myr, for all spaxels of the AGN and control samples obeying the criteria of Sec.\,\ref{validation}. The red line is the linear relation of a robust fit between log(\sfrgd) and log(\sfrsd) given inside the panels. We also list the number of spaxels and the Spearman's correlation coefficient ($r$) of the relation. For more details see text.}
    \label{allbins}
\end{figure*}

\begin{figure*}
    \centering
     \includegraphics[width=0.95\textwidth,height=0.95\textheight, keepaspectratio]{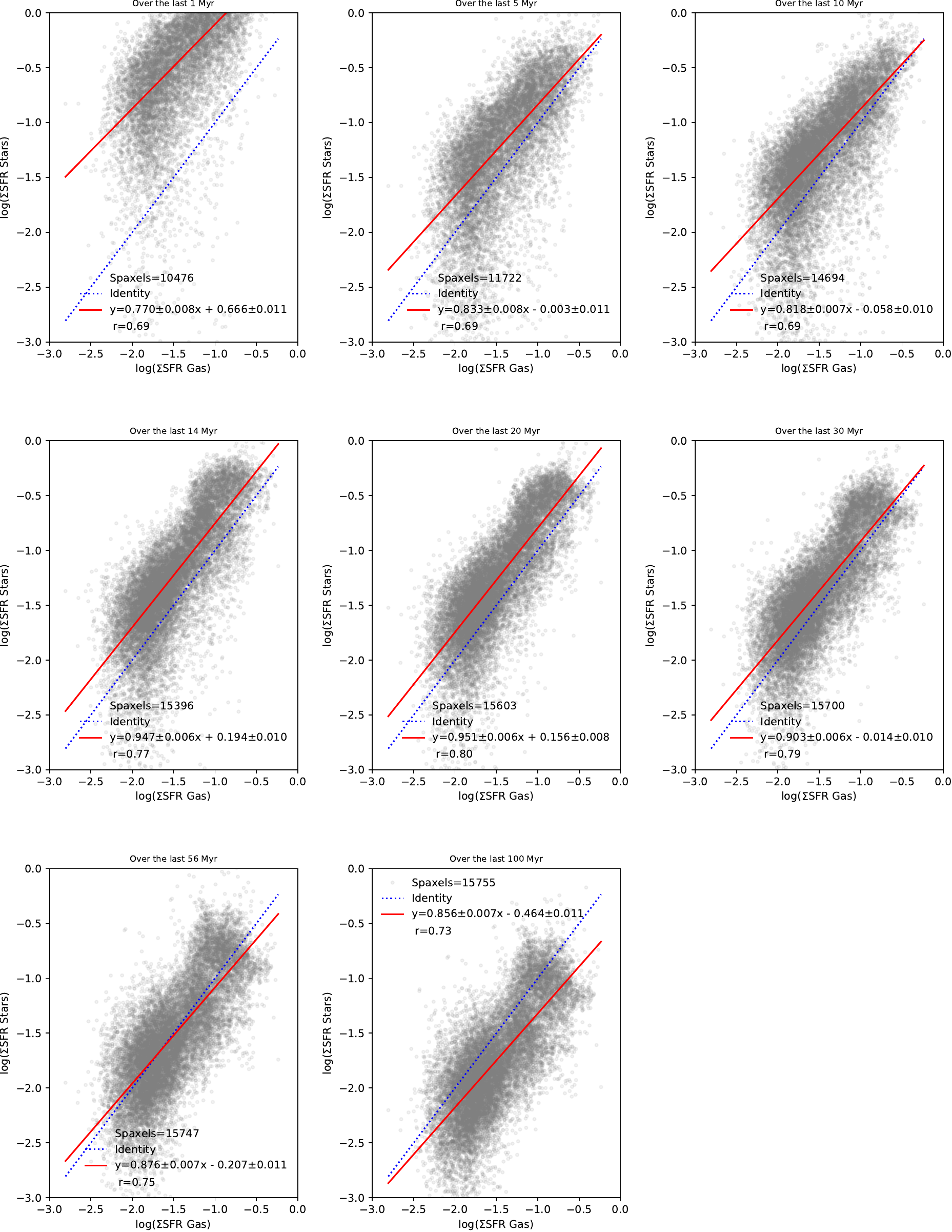}
    \caption{Same as Fig.~\ref{allbins} but only for the star-forming spaxels in the AGN sample.}
    \label{allbinsAGN}
\end{figure*}

The \sfrsd\ values were calculated over different age bins, comprising values over the last 1,  5, 10, 14, 20, 30, 56 and 100~Myr, each age bin corresponding to the added contribution of all younger age bins. These \sfrsd\ values are compared to the \sfrgd\ ones in Fig.~\ref{allbins} for the total sample (sum of AGN and control samples) and in Fig.~\ref{allbinsAGN} for the SF spaxels only in the AGN hosts. We also present the identity line (e.g. x=y; dotted blue) together with a linear fit to the data points (solid red line) obtained using bootstrap realisations \citep{Davison+97} with Huber Regressor model that is robust to outliers \citep{Owen+07}. The Spearman's correlation coefficient ($r$) and the number of spaxels are also included in the panels showing the plots. 

What emerges from this exercise is that the best correlation ($r=0.62$ for SF of the control+AGN samples and $r=0.80$ if only SF spaxels in the AGN sample are considered) is obtained when comparing the  \sfrsd\ over the last 20~Myr with that obtained with \sfrgd\ (from the \ha\ emission line). Also, their values are close to a one to one relation, with $\mbox{log(\sfrsd)}= (0.79\pm0.006)\,\mbox{log(\sfrgd)}-(0.22\pm0.009)$. This is also the stellar population age range that shows the smallest scatter of the points compared with the \sfrsd\ derived over the other $\Delta t$'s. This result is not surprising since the good agreement between the \sfrgd\ and  \sfrsd\ over the last 20~Myr is related to the fact that the stars which dominate the total ionising photons budget are the hot, massive ($M >10 M_\odot$) and short-lived ($t < 20~ \mbox{Myr}$) stars. Thus, as the emission-line fluxes provide an `instantaneous' measure of the SFR \citep{KennicuttJr.+98}, the corresponding SFR values should be more similar to those obtained from recent $\rm{SFR}_\star$. In fact, our results are in agreement with the previous findings of \citet{Asari+07} who found that the $\rm{SFR}_\star$ of the last 25\,Myr correlates well with that derived via nebular emission when using SDSS single  fibber observations. 

The above finding shows that one can 
use the \sfrsd\ over the last 20~Myr as a probe of the recent, instantaneous SFR that a galaxy is experiencing. 

\begin{figure*}
    \centering
     \includegraphics[width=\textwidth,height=\textheight, keepaspectratio]{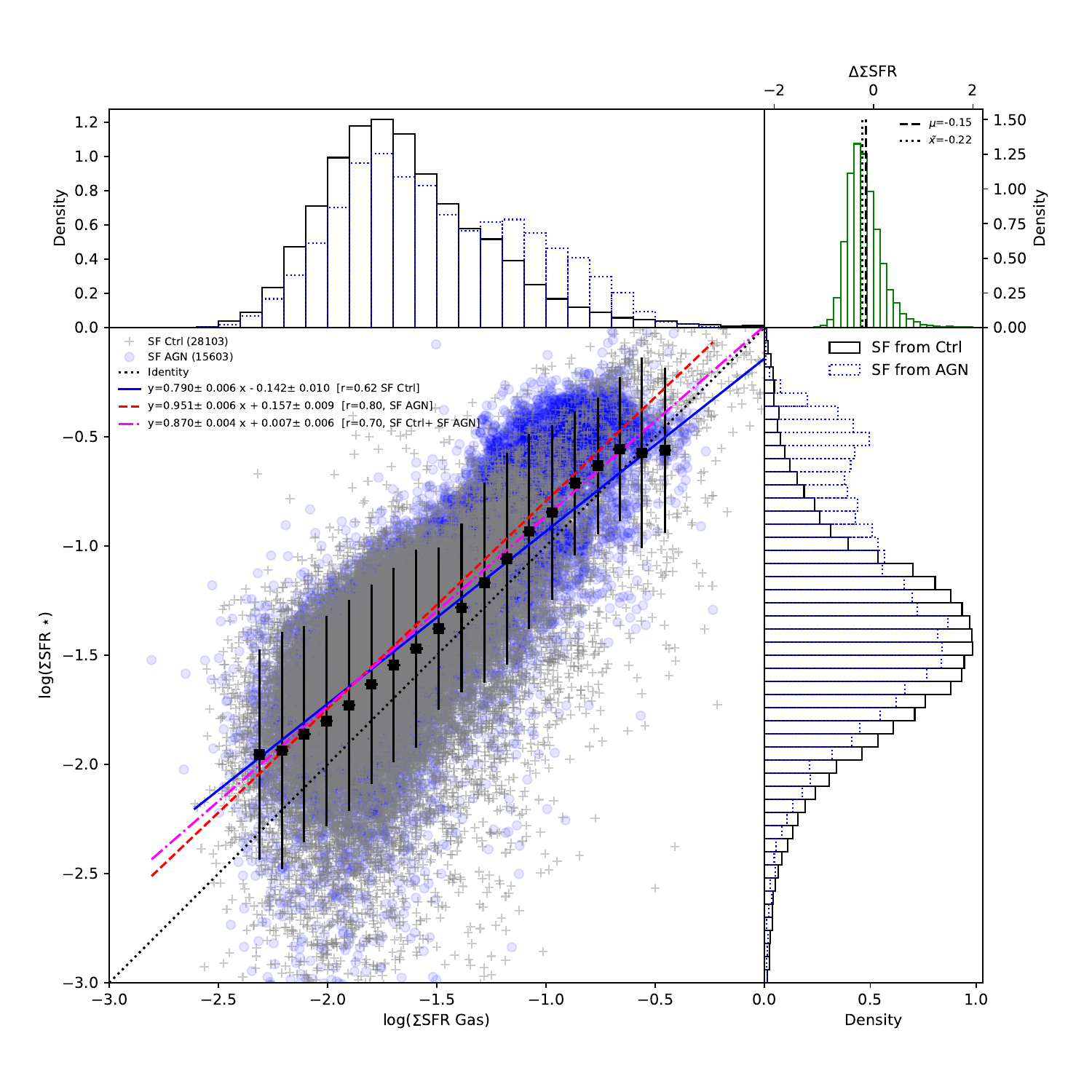}
     \caption{\sfrsd\  {\it versus}  \sfrgd\ over the last 20~Myrs in log scale. Spaxels with SF line ratios taken from the control sample (``Ctrl'') are represented as grey plus symbols, SF spaxels taken from AGN hosts as light blue circles. Regressions for each data set are labelled. Squares represent the mean value with standard deviations of 20 linearly spaced bins over log(\sfrgd) (values below 0.5 and 99.5 of the q-th percentile were removed) considering all spaxels (SF Ctrl and SF AGN). Density histograms for \sfrgd\ and \sfrsd\ of both samples are also shown. The top right histogram shows the $\Delta\Sigma = \mbox{log(\sfrgd)} - \mbox{log(\sfrsd)}$ as well as the mean ($\mu$) and median ($\tilde x$) values of this difference.}
    \label{SFR20}
\end{figure*}

In order to investigate any possible difference in behaviour between the spaxels from the AGN host galaxies and those from the control galaxies, we have plotted in Fig.~\ref{SFR20} the \sfrgd\ {\it versus} \sfrsd\ over the last 20\,Myr (for spaxels with SF line ratios) as grey plus symbols for the control sample and as light blue circles for the AGN hosts. We also plot the mean values, with standard deviations,  considering all the spaxels (both from control and AGN galaxies) divided in 20 linearly spaced bins over \sfrgd. Values below the 0.5th percentile and above the 99.5th percentile were removed to better display the results. The identity line (dotted black) is shown, as well as separate regressions for the two samples (AGN and control) and for their combined sample. We also show histograms for the density distribution\footnote{The counts are normalised to form a probability density, i.e. the integral under the histogram is 1. This is achieved by dividing the count by the number of observations times the bin width and not dividing by the total number of observations. The density histograms were obtained setting \texttt{density=True} in Python's {\sc matplotlib.pyplot.hist} routine.} of \sfrgd\ and \sfrsd\ for both samples, as well as an histogram showing the $\Delta\Sigma = \mbox{log(\sfrgd)} - \mbox{log(\sfrsd)}$ together with the mean ($\mu$) and median ($\tilde x$) values of this difference.

Besides obtaining the linear regressions to the data, as described above, we derived the Spearman's correlation coefficients, that are also listed within the Figs.~1 and 3 panels. These figures show that: (i) when considering only SF spaxels from the control sample we found $\mbox{log(\sfrsd)} = (0.790\pm0.006)\mbox{log(\sfrgd)} - (0.142\pm0.010)$ with $r=0.62$; (ii) when using the SF spaxels from the AGN hosts we found  $\mbox{log(\sfrsd)} = (0.951\pm0.006)\mbox{log(\sfrgd)} + (0.157\pm0.009)$ with $r=0.80$; and (iii) finally when combining both samples we find $\mbox{log(\sfrsd)} = (0.870\pm0.004)\mbox{log(\sfrgd)} + (0.007\pm0.006)$ with $r=0.70$. The three best-fit lines in Fig.~3 show a very similar slope, and a nearly one to one correlation is found.
The difference histogram ($\Delta\Sigma$SFR) confirms that both $\Sigma$SFRs are very similar, with the bulk of the differences being concentrated around $\sim$0. This result suggests that the \sfrsd over the last 20~Myrs can be directly used as a measure of the \sfrgd, for both star forming and AGN hosts (when the  spaxel line ratios indicate SF excitation). 

At the highest \sfrgd\ and \sfrsd\ values, there is a clear excess of AGN SF spaxels relative to those from the control ones (histograms in Fig.~\ref{SFR20}). In order to understand the origin of this excess, our first step was to remove the strong AGNs, defined as the sources with $L(\rm{[OIII]}\lambda 5007\AA) \ge 3.8\times 10^{40}\,\mbox{erg}\,\mbox{s}^{-1} $  \citep[][ ]{Rembold+17,Mallmann+18a} from our sample, since, if both the AGN activity and the SF would be driven by the same mechanism (e.g. larger gas reservoirs both forming stars and feeding the AGN) this could explain this excess. However, after the removal of these objects the same trend remains and the excess is still observed.  In order to identify the galaxies responsible for the excess, we applied a cut for high  $\Sigma$SFRs. Selecting only spaxels with $\log(\mbox{\sfrgd}) >  -1.0$ and $\log(\mbox{\sfrsd}) > -0.75$, we found that all the spaxels presenting these high values come from only four AGN hosts identified by the following MaNGA-IDs: 1-189584, 1-604022, 1-258373,1-229731 (Figs.~\ref{Ap1} -- ~\ref{Ap4}). Of these, only objects 1-258373 and 1-229731 are strong AGNs. This result shows that not only strong AGNs are producing the high $\Sigma$SFRs in Fig.~\ref{SFR20}.

\begin{figure*}
    \centering
     \includegraphics[width=\textwidth,height=\textheight, keepaspectratio]{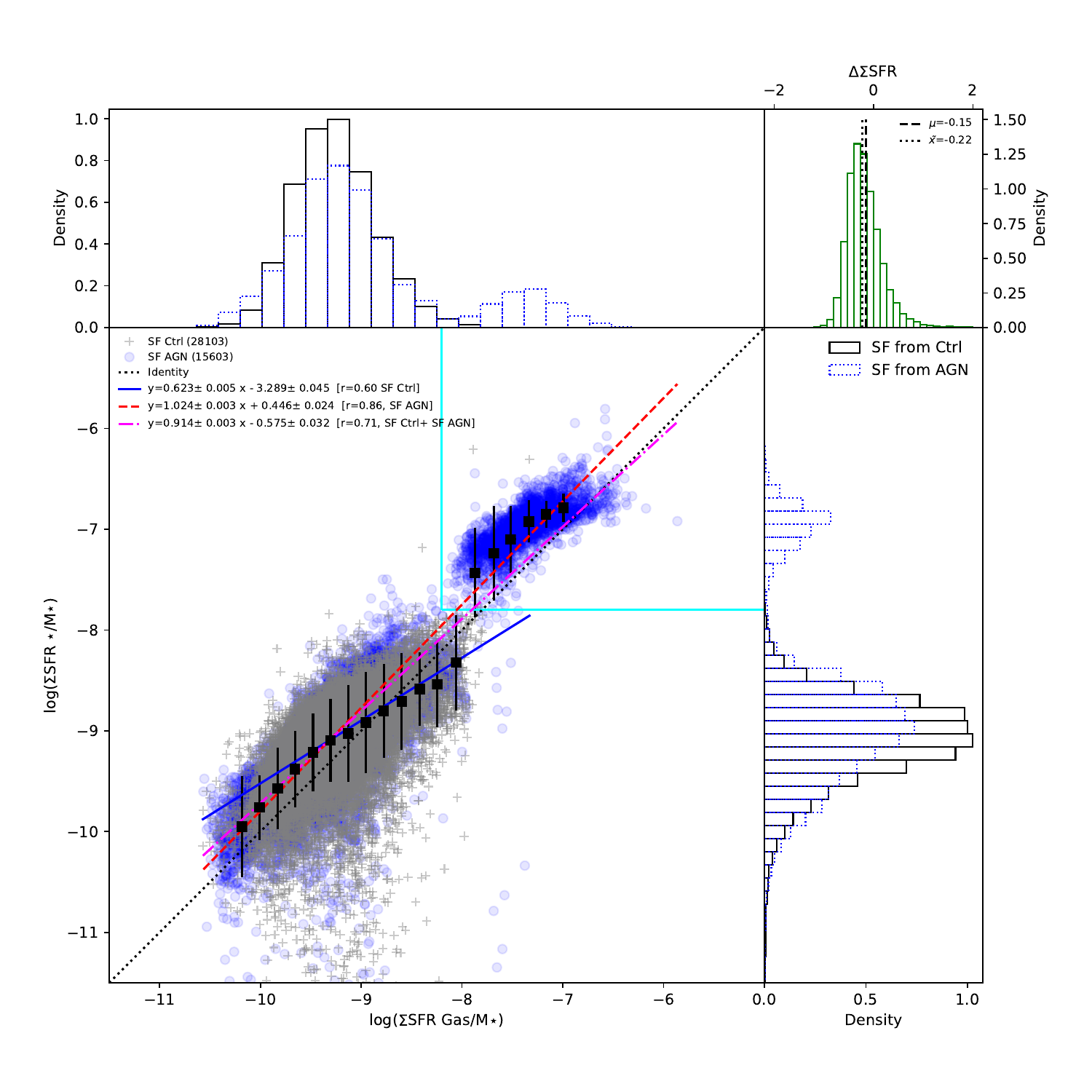}
     \caption{\sfrgd\  {\it versus}  \sfrsd over the last 20~Myrs normalized by stellar mass. Spaxels with SF line ratios taken from the control sample are grey plus symbols, SF spaxels taken from AGN hosts are light blue circles. Regressions over each data set are labelled. Squares represent the mean value with standard deviations  of 20 linearly spaced bins over \sfrgd (values below 0.5 and 99.5 of the  q-th percentile were removed) considering all spaxels (SF from Ctrl and SF AGNs). Density histograms of both samples are also shown. The cyan rectangle represents the high $\Sigma\mbox{SFR}/M_\star$ region, namely: $\log(\Sigma\mbox{SFR}_{g}/M_{\star}) > -8.2$ and $\log(\Sigma \mbox{SFR}_{\star}/M_{\star}) > -7.8$.}
    \label{SFR20Mass}
\end{figure*}

Since a tight correlation between the SFR and the stellar mass of galaxies is expected, and indeed observed in the form of the so-called \emph{star formation Main Sequence} (MS) of galaxies \citep[][it is also observed when considering individual spaxels, e.g.  \citealt{Lin+19}]{Brinchmann+04,Noeske+07,Daddi+07,Speagle+14}, we decided to normalise the $\Sigma$SFRs to the stellar mass, $M_\star$, of each spaxel. The result of this normalisation is shown in Fig.~\ref{SFR20Mass}, where one clearly sees that the spaxels contributing to the tail observed in the distributions of the AGN spaxels in the previous histograms are those with the highest $\Sigma$SFR/M$_\star$, and they clearly populate a separate region in this figure.

In order to investigate the origin of these high $\Sigma\mbox{SFR}/M_\star$'s, we have tracked them back to the sources originating such spaxels. Therefore, using  Fig.~\ref{SFR20Mass} we defined  the limits of the high-$\Sigma\mbox{SFR}/M_\star$ ``cloud'' as  $\log(\mbox{\sfrgd}/M_{\star}) > -8.2$ and $\log(\mbox{\sfrsd}/M_{\star}) > -7.8$ (cyan rectangle). The spaxels located in this region are from four AGNs, identified by the  MANGAIDs: 1-189584, 1-604022 (these two being also identified as producing the high values in Figs.\,\ref{allbins} and \ref{SFR20}, plus 1-603941, 1-420924, none of which is classified as strong AGNs. 

This result suggests that these sources are somewhat particular in the sense of having a significantly larger star-forming gas reservoir leading to a higher star-formation rate than the other sources in our sample. Interestingly, all these sources present companions or satellite galaxies at projected distances smaller than 70\,kpc. In addition, objects 1-189584, 1-604022 and 1-603941 are members of galaxy groups \citep{Fouque+92,White+99,VonDerLinden+07}. This suggests that the higher star formation efficiencies shown by these objects relative to other galaxies of similar masses is due to interactions with nearby galaxies. A definitive answer to this question requires a complete statistics of the environment of all galaxies in the sample, which is beyond the scope of this work but will be addressed in a forthcoming publication.

The results presented here clearly show that we can use the transformation equation:

\begin{equation}
\mbox{log(\sfrsd)} = (0.870\pm0.004)\mbox{log(\sfrgd)} +(0.007\pm0.006)
\label{t_eq} \\
\end{equation}
or
\begin{equation*}
\mbox{log(\sfrgd)} = (1.149\pm0.005)\mbox{log(\sfrsd)}  - (0.008\pm0.008)
\end{equation*}
 to obtain the gas $\Sigma$SFR from the stellar one.

 The above result is particularly useful for obtaining SFR values in the narrow-line region (NLR) or extended NLR (ENLR) of AGN hosts. Since the synthesis technique allows to fit the stellar populations disentangling it from the AGN featureless continuum, it is possible to use it to obtain the SFRs that one would derive from the gas emission, even in regions dominated by the AGN excitation, such as the NLR and ENLR, for which HI emission cannot be used as a star-formation indicator. 
 
 These results can thus be used to investigate if there is, for example, SF quenching (or SF enhancement) in the vicinity of AGN, comparing the results with the predictions of AGN feedback effects in cosmological simulations \citep[e.g.][]{Nelson+15,McAlpine+16}. In a forthcoming publication we intend to apply the relation of eq.\,\ref{t_eq} in order to compare the $\Sigma$SFR we obtain in regions dominated by AGNs excitation to those obtained at similar distances from the nucleus for control galaxies in order to investigate any difference related to the AGN. The results will then be compared to those of cosmological simulations predictions  (Schimoia et al. {\it in preparation}). In addition, we will extend the investigation previously done for a smaller sample \citep{Mallmann+18a} on the systematic stellar population differences between AGN hosts and controls for the updated sample used here (Mallmann et al. {\it in preparation}).

\subsection{Comparing the nebular and stellar population reddening}

\begin{figure*}
    \centering
     \includegraphics[width=\textwidth,height=\textheight, keepaspectratio]{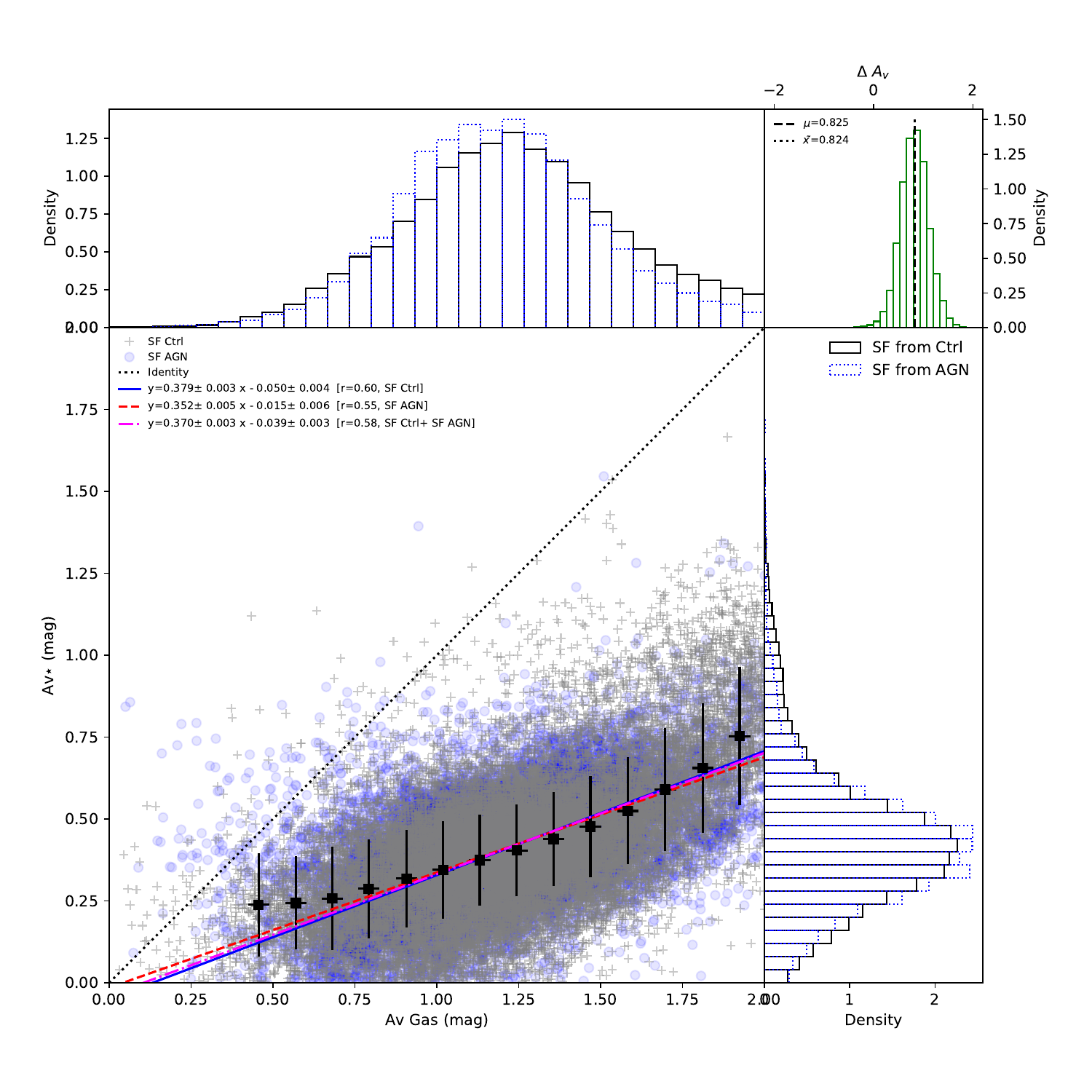}
    \caption{Gas reddening {\it versus} full spectral fitting reddening. Spaxels with SF line ratios taken from the control sample are grey plus symbols, SF spaxels taken from AGN hosts are light blue circles. Regressions over each data set are labelled. Squares represent the mean value with standard deviations  of 20 linearly spaced bins over \sfrgd (values below 0.5 and 99.5 of the  q-th percentile were removed) considering all spaxels (SF from Ctrl and SF AGNs). Density histograms of both samples are also shown. }
    \label{AV}
\end{figure*}

A comparison between the $A_V$ values obtained via \ha/\hb\ line ratios and that from the full spectral fitting of the stellar population is shown in Fig.~\ref{AV}.
We fitted linear regressions to the data and derived the Spearman's correlation coefficients. We found that: (i) when considering only SF spaxels from the control sample  $A_{V,\star} = (0.379\pm0.003)A_{V,g} - (0.050\pm0.004)$, with $r=0.60$; (ii) when using the SF spaxels from the AGN hosts,
 $A_{V,\star} = (0.352\pm0.005)A_{V,g} - (0.015\pm0.006)$, with $r=0.55$; and (iii) when considering the combined samples of AGN and controls we find  $A_{V,\star} = (0.370\pm0.003)A_{V,g} - (0.039\pm0.003)$, with $r=0.58$.

This result suggests that the reddening derived via stellar population full spectral fitting is consistent for both samples and that the extinction derived for the gas $A_{V,g}$ is consistently larger than $A_{V,\star}$ by a multiplicative factor ranging from $2.64$ to $2.85$, depending on the sample. The larger values obtained for the gas extinction than those of the stars is in agreement with the similar finding of \citet{Calzetti+94}, who have analysed IUE UV and optical spectra of 39 starburst and blue compact galaxies and studied the average properties of dust extinction. They found that the optical depth obtained via Balmer emission lines is about twice that obtained via the underlying continuum around these lines.
They interpret this difference as a consequence of the fact that the hot ionising stars are associated with dustier regions than that of older (colder) stars that contribute also to the continuum. 

We thus interpret the difference we found for the extinction of the stellar population via full spectral fitting and that of the gas as being due to the fact that \str\ works with a single reddening for all the population components (e.g. the final mix is reddened). Since the old component contributes significantly ($\gtrsim$ 60 per cent) to the integrated light in almost all spaxels \citep[see Fig.1 of][]{Mallmann+18a}, its reddening dominates the reddening of the final integrated model spectrum. The reddening of this older and cooler underlying population is therefore lower than that of the younger components, that are buried deeper in the dust (and are responsible to ionise  the line-emitting gas).


\section{Conclusions } \label{sec:conclusions}

We have presented a comparison between star-formation rate surface densities obtained from spectral synthesis of the stellar population \sfrsd\ and from the H$\alpha$ gas emission \sfrgd\ for an updated sample of 170 AGN and 291 control galaxies relative to our initial sample defined by \citet{Rembold+17}.  We have used the corresponding MaNGA datacubes selecting only spaxels high SNR continuum, with strong emission lines and showing star-forming line ratios as obtained from diagnostic diagrams. Our main results can be summarised as follows.

\begin{itemize}
    \item The \sfrsd\ over the last 20\,Myrs and \sfrgd\ shows the best correlation among all tested age bins, both including in the analysis the SF spaxels from AGN hosts or only those from control galaxies. The transformation equation is $\mbox{log(\sfrsd)} = (0.870\pm0.004)\mbox{log(\sfrgd)} +(0.007\pm0.006)$ or $\mbox{log(\sfrgd)} = (1.149\pm0.005)\mbox{log(\sfrsd)}  - (0.008\pm0.008)$. This result opens a new way to obtain the SFRs in AGN hosts, even in the NLR and ENLR, were the AGN dominates the excitation of the emission lines and the SFR cannot be obtained directly from the HI line fluxes.
    
    \item A few AGN hosts show an excess of $\Sigma$SFR relative to the rest of the sample which we tentatively attribute to a larger gas reservoir and star formation efficiency. Coincidentally, these AGNs seem to have close neighbours, thus the interaction could have boosted the star formation (to be further investigated due to the small size of the sample).
    
    \item We found that the visual extinction $A_{V,g}$ derived from the Balmer decrement is 2.63 to 2.86 times larger than the extinction derived from the stellar population synthesis, $A_{V,star}$. This result is in agreement with previous literature results based on much smaller samples. We interpret the difference as being due to the fact that \str\ works with a single reddening for all populations, and the reddening of the stellar content is dominated by the older population that is less extincted than the young stellar population that would be similarly extincted as the star-forming gas. 
    
\end{itemize}

The transformation equation presented here (Eq.~\ref{t_eq}) can be used to obtain the gas $\Sigma$SFR in AGN hosts via stellar population synthesis using the full spectral fitting. Since the synthesis allows disentangling the contributions of the stellar populations from that of the AGN featureless continuum, it is possible to obtain the $\Sigma$SFRs that one would derive for the gas emission, even in regions dominated by AGN excitation. The obtained SFRs in AGN hosts can then be compared with those obtained for control galaxies to investigate the effect of AGN on the surrounding stellar population as well as be compared (and incorporated) with the SFR values predicted as due to AGN feedback effects on the host galaxies in cosmological simulations.

\section*{Acknowledgements}

We thank an anonymous referee for comments and suggestions that have helped improving the text. R.R. Thanks CNPq, CAPES and FAPERGS for financial support, as well as to Marina Trevisan for useful discussions of the present results.
RAR thanks partial financial support from Conselho Nacional de Desenvolvimento Cient\'ifico e Tecnol\'ogico (202582/2018-3 and 302280/2019-7) and Funda\c c\~ao de Amparo \`a pesquisa do Estado do Rio Grande do Sul (17/2551-0001144-9 and 16/2551-0000251-7).
MB acknowledges support from FONDECYT regular grant 1170618.
GSC acknowledges the support from CONICYT FONDECYT project No. 3190561.

SDSS is managed by the Astrophysical Research Consortium for the Participating Institutions of the SDSS Collaboration including the Brazilian Participation Group, the Carnegie Institution for Science, Carnegie Mellon University, the Chilean Participation Group, the French Participation Group, Harvard-Smithsonian Center for Astrophysics, Instituto de Astrofisica de Canarias, The Johns Hopkins University, Kavli Institute for the Physics and Mathematics of the Universe (IPMU) / University of Tokyo, the Korean Participation Group, Lawrence Berkeley National Laboratory, Leibniz Institut f\"ur Astrophysik Potsdam (AIP), Max-Planck-Institut f\"ur Astronomie (MPIA Heidelberg), Max-Planck-Institut f\"ur Astrophysik (MPA Garching), Max-Planck-Institut f\"ur Extraterrestrische Physik (MPE), National Astronomical Observatories of China, New Mexico State University, New York University, University of Notre Dame, Observat\' orio Nacional / MCTI, The Ohio State University, Pennsylvania State University, Shanghai Astronomical Observatory, United Kingdom Participation Group, Universidad Nacional Aut\'onoma de M\'exico, University of Arizona, University of Colorado Boulder, University of Oxford, University of Portsmouth, University of Utah, University of Virginia, University of Washington, University of Wisconsin, Vanderbilt University, and Yale University.

This research made use of Astropy,\footnote{http://www.astropy.org} a community-developed core Python package for Astronomy \citep{AstropyCollaboration+13,AstropyCollaboration+18}. 

\section*{Data Availability}

 The data underlying this article are available under SDSS collaboration rules, and the by products will be shared on reasonable request to the corresponding author.

\bibliographystyle{mnras}
\bibliography{ms_riffel_Rev1.bib}

\begin{thebibliography}{}
\makeatletter
\relax
\def\mn@urlcharsother{\let\do\@makeother \do\$\do\&\do\#\do\^\do\_\do\%\do\~}
\def\mn@doi{\begingroup\mn@urlcharsother \@ifnextchar [ {\mn@doi@}
  {\mn@doi@[]}}
\def\mn@doi@[#1]#2{\def\@tempa{#1}\ifx\@tempa\@empty \href
  {http://dx.doi.org/#2} {doi:#2}\else \href {http://dx.doi.org/#2} {#1}\fi
  \endgroup}
\def\mn@eprint#1#2{\mn@eprint@#1:#2::\@nil}
\def\mn@eprint@arXiv#1{\href {http://arxiv.org/abs/#1} {{\tt arXiv:#1}}}
\def\mn@eprint@dblp#1{\href {http://dblp.uni-trier.de/rec/bibtex/#1.xml}
  {dblp:#1}}
\def\mn@eprint@#1:#2:#3:#4\@nil{\def\@tempa {#1}\def\@tempb {#2}\def\@tempc
  {#3}\ifx \@tempc \@empty \let \@tempc \@tempb \let \@tempb \@tempa \fi \ifx
  \@tempb \@empty \def\@tempb {arXiv}\fi \@ifundefined
  {mn@eprint@\@tempb}{\@tempb:\@tempc}{\expandafter \expandafter \csname
  mn@eprint@\@tempb\endcsname \expandafter{\@tempc}}}

\bibitem[\protect\citeauthoryear{Aghanim et~al.,}{Aghanim
  et~al.}{2020}]{Aghanim+20}
Aghanim N.,  et~al., 2020, \mn@doi [Astronomy \& Astrophysics]
  {10.1051/0004-6361/201833910}, 641, A6

\bibitem[\protect\citeauthoryear{Aguado et~al.,}{Aguado
  et~al.}{2019}]{Aguado+19}
Aguado D.~S.,  et~al., 2019, \mn@doi [The Astrophysical Journal Supplement
  Series] {10.3847/1538-4365/aaf651}, 240, 23

\bibitem[\protect\citeauthoryear{Asari, Cid~Fernandes, Stasi{\'n}ska,
  {Torres-Papaqui}, Mateus, Sodr{\'e}, Schoenell  \& Gomes}{Asari
  et~al.}{2007}]{Asari+07}
Asari N.~V.,  Cid~Fernandes R.,  Stasi{\'n}ska G.,  {Torres-Papaqui} J.~P.,
  Mateus A.,  Sodr{\'e} L.,  Schoenell W.,   Gomes J.~M.,  2007, \mn@doi
  [Monthly Notices of the Royal Astronomical Society]
  {10.1111/j.1365-2966.2007.12255.x}, 381, 263

\bibitem[\protect\citeauthoryear{{Astropy Collaboration} et~al.,}{{Astropy
  Collaboration} et~al.}{2013}]{AstropyCollaboration+13}
{Astropy Collaboration} et~al., 2013, \mn@doi [Astronomy and Astrophysics]
  {10.1051/0004-6361/201322068}, 558, A33

\bibitem[\protect\citeauthoryear{{Astropy Collaboration} et~al.,}{{Astropy
  Collaboration} et~al.}{2018}]{AstropyCollaboration+18}
{Astropy Collaboration} et~al., 2018, \mn@doi [The Astronomical Journal]
  {10.3847/1538-3881/aabc4f}, 156, 123

\bibitem[\protect\citeauthoryear{Baldry, Glazebrook, Brinkmann, Ivezi{\'c},
  Lupton, Nichol  \& Szalay}{Baldry et~al.}{2004}]{Baldry+04}
Baldry I.~K.,  Glazebrook K.,  Brinkmann J.,  Ivezi{\'c} {\v Z}.,  Lupton
  R.~H.,  Nichol R.~C.,   Szalay A.~S.,  2004, \mn@doi [The Astrophysical
  Journal] {10.1086/380092}, 600, 681

\bibitem[\protect\citeauthoryear{Baldwin, McDermid, Kuntschner, Maraston  \&
  Conroy}{Baldwin et~al.}{2018}]{Baldwin+18}
Baldwin C.,  McDermid R.~M.,  Kuntschner H.,  Maraston C.,   Conroy C.,  2018,
  \mn@doi [Monthly Notices of the Royal Astronomical Society]
  {10.1093/mnras/stx2502}, 473, 4698

\bibitem[\protect\citeauthoryear{{Barrera-Ballesteros}
  et~al.,}{{Barrera-Ballesteros} et~al.}{2020}]{Barrera-Ballesteros+20}
{Barrera-Ballesteros} J.~K.,  et~al., 2020, \mn@doi [Monthly Notices of the
  Royal Astronomical Society] {10.1093/mnras/stz3553}, 492, 2651

\bibitem[\protect\citeauthoryear{Belfiore et~al.,}{Belfiore
  et~al.}{2019}]{Belfiore+19}
Belfiore F.,  et~al., 2019, \mn@doi [The Astronomical Journal]
  {10.3847/1538-3881/ab3e4e}, 158, 160

\bibitem[\protect\citeauthoryear{Bieri, Dubois, Silk, Mamon  \& Gaibler}{Bieri
  et~al.}{2016}]{Bieri+16}
Bieri R.,  Dubois Y.,  Silk J.,  Mamon G.~A.,   Gaibler V.,  2016, \mn@doi
  [Monthly Notices of the Royal Astronomical Society] {10.1093/mnras/stv2551},
  455, 4166

\bibitem[\protect\citeauthoryear{Blanton et~al.,}{Blanton
  et~al.}{2017}]{Blanton+17}
Blanton M.~R.,  et~al., 2017, \mn@doi [The Astronomical Journal]
  {10.3847/1538-3881/aa7567}, 154, 28

\bibitem[\protect\citeauthoryear{Brammer et~al.,}{Brammer
  et~al.}{2009}]{Brammer+09}
Brammer G.~B.,  et~al., 2009, \mn@doi [The Astrophysical Journal]
  {10.1088/0004-637X/706/1/L173}, 706, L173

\bibitem[\protect\citeauthoryear{Brinchmann, Charlot, White, Tremonti,
  Kauffmann, Heckman  \& Brinkmann}{Brinchmann et~al.}{2004}]{Brinchmann+04}
Brinchmann J.,  Charlot S.,  White S. D.~M.,  Tremonti C.,  Kauffmann G.,
  Heckman T.,   Brinkmann J.,  2004, \mn@doi [Monthly Notices of the Royal
  Astronomical Society] {10.1111/j.1365-2966.2004.07881.x}, 351, 1151

\bibitem[\protect\citeauthoryear{Bruzual \& Charlot}{Bruzual \&
  Charlot}{2003}]{Bruzual+03a}
Bruzual G.,  Charlot S.,  2003, \mn@doi [Monthly Notices of the RAS]
  {10.1046/j.1365-8711.2003.06897.x}, \href
  {http://adsabs.harvard.edu/abs/2003MNRAS.344.1000B} {344, 1000}

\bibitem[\protect\citeauthoryear{Bundy et~al.,}{Bundy et~al.}{2015}]{Bundy+15a}
Bundy K.,  et~al., 2015, \mn@doi [Astrophysical Journal]
  {10.1088/0004-637X/798/1/7}, \href
  {http://adsabs.harvard.edu/abs/2015ApJ...798....7B} {798, 7}

\bibitem[\protect\citeauthoryear{Calzetti, Kinney  \&
  {Storchi-Bergmann}}{Calzetti et~al.}{1994}]{Calzetti+94}
Calzetti D.,  Kinney A.~L.,   {Storchi-Bergmann} T.,  1994, \mn@doi [The
  Astrophysical Journal] {10.1086/174346}, 429, 582

\bibitem[\protect\citeauthoryear{Calzetti, Armus, Bohlin, Kinney, Koornneef  \&
  {Storchi-Bergmann}}{Calzetti et~al.}{2000}]{Calzetti+00}
Calzetti D.,  Armus L.,  Bohlin R.~C.,  Kinney A.~L.,  Koornneef J.,
  {Storchi-Bergmann} T.,  2000, \mn@doi [The Astrophysical Journal]
  {10.1086/308692}, 533, 682

\bibitem[\protect\citeauthoryear{Cardelli, Clayton  \& Mathis}{Cardelli
  et~al.}{1989}]{Cardelli+89a}
Cardelli J.~A.,  Clayton G.~C.,   Mathis J.~S.,  1989, \mn@doi [Astrophysical
  Journal] {10.1086/167900}, \href
  {http://adsabs.harvard.edu/abs/1989ApJ...345..245C} {345, 245}

\bibitem[\protect\citeauthoryear{Cherinka et~al.,}{Cherinka
  et~al.}{2019}]{Cherinka+19}
Cherinka B.,  et~al., 2019, \mn@doi [The Astronomical Journal]
  {10.3847/1538-3881/ab2634}, 158, 74

\bibitem[\protect\citeauthoryear{Cid~Fernandes}{Cid~Fernandes}{2018}]{CidFernandes+18}
Cid~Fernandes R.,  2018, \mn@doi [Monthly Notices of the Royal Astronomical
  Society] {10.1093/mnras/sty2012}, 480, 4480

\bibitem[\protect\citeauthoryear{Cid~Fernandes, Gu, Melnick, Terlevich,
  Terlevich, Kunth, Rodrigues~Lacerda  \& Joguet}{Cid~Fernandes
  et~al.}{2004}]{CidFernandes+04}
Cid~Fernandes R.,  Gu Q.,  Melnick J.,  Terlevich E.,  Terlevich R.,  Kunth D.,
   Rodrigues~Lacerda R.,   Joguet B.,  2004, \mn@doi [Monthly Notices of the
  RAS] {10.1111/j.1365-2966.2004.08321.x}, \href
  {http://adsabs.harvard.edu/abs/2004MNRAS.355..273C} {355, 273}

\bibitem[\protect\citeauthoryear{Cid~Fernandes, Mateus, Sodr{\'e},
  Stasi{\'n}ska  \& Gomes}{Cid~Fernandes et~al.}{2005}]{CidFernandes+05}
Cid~Fernandes R.,  Mateus A.,  Sodr{\'e} L.,  Stasi{\'n}ska G.,   Gomes J.~M.,
  2005, \mn@doi [Monthly Notices of the RAS]
  {10.1111/j.1365-2966.2005.08752.x}, \href
  {http://adsabs.harvard.edu/abs/2005MNRAS.358..363C} {358, 363}

\bibitem[\protect\citeauthoryear{Cid~Fernandes et~al.,}{Cid~Fernandes
  et~al.}{2013}]{CidFernandes+13}
Cid~Fernandes R.,  et~al., 2013, \mn@doi [Astronomy and Astrophysics]
  {10.1051/0004-6361/201220616}, 557, A86

\bibitem[\protect\citeauthoryear{Cid~Fernandes et~al.,}{Cid~Fernandes
  et~al.}{2014}]{CidFernandes+14}
Cid~Fernandes R.,  et~al., 2014, \mn@doi [Astronomy and Astrophysics]
  {10.1051/0004-6361/201321692}, 561, A130

\bibitem[\protect\citeauthoryear{Crain et~al.,}{Crain et~al.}{2015}]{Crain+15}
Crain R.~A.,  et~al., 2015, \mn@doi [Monthly Notices of The Royal Astronomical
  Society] {10.1093/mnras/stv725}, 450, 1937

\bibitem[\protect\citeauthoryear{Croton et~al.,}{Croton
  et~al.}{2006}]{Croton+06}
Croton D.~J.,  et~al., 2006, \mn@doi [Monthly Notices of The Royal Astronomical
  Society] {10.1111/j.1365-2966.2005.09675.x}, 365, 11

\bibitem[\protect\citeauthoryear{Daddi et~al.,}{Daddi et~al.}{2007}]{Daddi+07}
Daddi E.,  et~al., 2007, \mn@doi [The Astrophysical Journal] {10.1086/521818},
  670, 156

\bibitem[\protect\citeauthoryear{Davison \& Hinkley}{Davison \&
  Hinkley}{1997}]{Davison+97}
Davison A.~C.,  Hinkley D.~V.,  1997, Bootstrap {{Methods}} and Their
  {{Application}},
  /core/books/bootstrap-methods-and-their-application/ED2FD043579F27952363566DC09CBD6A,
  \mn@doi{10.1017/CBO9780511802843}

\bibitem[\protect\citeauthoryear{Dom{\'i}nguez et~al.,}{Dom{\'i}nguez
  et~al.}{2013}]{Dominguez+13}
Dom{\'i}nguez A.,  et~al., 2013, \mn@doi [The Astrophysical Journal]
  {10.1088/0004-637X/763/2/145}, 763, 145

\bibitem[\protect\citeauthoryear{Drory et~al.,}{Drory et~al.}{2015}]{Drory+15}
Drory N.,  et~al., 2015, \mn@doi [The Astronomical Journal]
  {10.1088/0004-6256/149/2/77}, 149, 77

\bibitem[\protect\citeauthoryear{{El-Badry}, Wetzel, Geha, Hopkins, Kere{\v s},
  Chan  \& {Faucher-Gigu{\`e}re}}{{El-Badry} et~al.}{2016}]{El-Badry+16}
{El-Badry} K.,  Wetzel A.,  Geha M.,  Hopkins P.~F.,  Kere{\v s} D.,  Chan
  T.~K.,   {Faucher-Gigu{\`e}re} C.-A.,  2016, \mn@doi [The Astrophysical
  Journal] {10.3847/0004-637X/820/2/131}, 820, 131

\bibitem[\protect\citeauthoryear{Fabian}{Fabian}{2012}]{Fabian+12}
Fabian A.~C.,  2012, \mn@doi [Annual Review of Astronomy and Astrophysics]
  {10.1146/annurev-astro-081811-125521}, 50, 455

\bibitem[\protect\citeauthoryear{Fouque, Gourgoulhon, Chamaraux  \&
  Paturel}{Fouque et~al.}{1992}]{Fouque+92}
Fouque P.,  Gourgoulhon E.,  Chamaraux P.,   Paturel G.,  1992, Astronomy and
  Astrophysics Supplement Series, 93, 211

\bibitem[\protect\citeauthoryear{Gallagher, Maiolino, Belfiore, Drory, Riffel
  \& Riffel}{Gallagher et~al.}{2019}]{Gallagher+19}
Gallagher R.,  Maiolino R.,  Belfiore F.,  Drory N.,  Riffel R.,   Riffel
  R.~A.,  2019, \mn@doi [Monthly Notices of the Royal Astronomical Society]
  {10.1093/mnras/stz564}, 485, 3409

\bibitem[\protect\citeauthoryear{Gonz{\'a}lez~Delgado, Cervi{\~n}o, Martins,
  Leitherer  \& Hauschildt}{Gonz{\'a}lez~Delgado
  et~al.}{2005}]{GonzalezDelgado+05}
Gonz{\'a}lez~Delgado R.~M.,  Cervi{\~n}o M.,  Martins L.~P.,  Leitherer C.,
  Hauschildt P.~H.,  2005, \mn@doi [Monthly Notices of the Royal Astronomical
  Society] {10.1111/j.1365-2966.2005.08692.x}, 357, 945

\bibitem[\protect\citeauthoryear{Granato, De~Zotti, Silva, Bressan  \&
  Danese}{Granato et~al.}{2004}]{Granato+04}
Granato G.~L.,  De~Zotti G.,  Silva L.,  Bressan A.,   Danese L.,  2004,
  \mn@doi [The Astrophysical Journal] {10.1086/379875}, 600, 580

\bibitem[\protect\citeauthoryear{Gunn et~al.,}{Gunn et~al.}{2006}]{Gunn+06}
Gunn J.~E.,  et~al., 2006, \mn@doi [The Astronomical Journal] {10.1086/500975},
  131, 2332

\bibitem[\protect\citeauthoryear{Hao, Kennicutt, Johnson, Calzetti, Dale  \&
  Moustakas}{Hao et~al.}{2011}]{Hao+11}
Hao C.-N.,  Kennicutt R.~C.,  Johnson B.~D.,  Calzetti D.,  Dale D.~A.,
  Moustakas J.,  2011, \mn@doi [The Astrophysical Journal]
  {10.1088/0004-637X/741/2/124}, 741, 124

\bibitem[\protect\citeauthoryear{Hopkins}{Hopkins}{2012}]{Hopkins+12}
Hopkins P.~F.,  2012, \mn@doi [Monthly Notices of the RAS]
  {10.1111/j.1745-3933.2011.01179.x}, \href
  {http://adsabs.harvard.edu/abs/2012MNRAS.420L...8H} {420, L8}

\bibitem[\protect\citeauthoryear{Ishibashi \& Fabian}{Ishibashi \&
  Fabian}{2012}]{Ishibashi+12}
Ishibashi W.,  Fabian A.~C.,  2012, \mn@doi [Monthly Notices of the Royal
  Astronomical Society] {10.1111/j.1365-2966.2012.22074.x}, 427, 2998

\bibitem[\protect\citeauthoryear{Kauffmann et~al.,}{Kauffmann
  et~al.}{2003a}]{Kauffmann+03a}
Kauffmann G.,  et~al., 2003a, \mn@doi [Monthly Notices of the Royal
  Astronomical Society] {10.1046/j.1365-8711.2003.06292.x}, 341, 54

\bibitem[\protect\citeauthoryear{Kauffmann et~al.,}{Kauffmann
  et~al.}{2003b}]{Kauffmann+03b}
Kauffmann G.,  et~al., 2003b, \mn@doi [Monthly Notices of the RAS]
  {10.1111/j.1365-2966.2003.07154.x}, \href
  {http://adsabs.harvard.edu/abs/2003MNRAS.346.1055K} {346, 1055}

\bibitem[\protect\citeauthoryear{Kennicutt}{Kennicutt}{1998}]{KennicuttJr.+98}
Kennicutt Jr. R.~C.,  1998, \mn@doi [Annual Review of Astron and Astrophys]
  {10.1146/annurev.astro.36.1.189}, \href
  {http://adsabs.harvard.edu/abs/1998ARA\%26A..36..189K} {36, 189}

\bibitem[\protect\citeauthoryear{Kennicutt \& Evans}{Kennicutt \&
  Evans}{2012}]{Kennicutt+12}
Kennicutt R.~C.,  Evans N.~J.,  2012, \mn@doi [Annual Review of Astronomy and
  Astrophysics] {10.1146/annurev-astro-081811-125610}, 50, 531

\bibitem[\protect\citeauthoryear{Kennicutt Jr. et~al.,}{Kennicutt
  et~al.}{2009}]{Kennicutt+09}
Kennicutt Jr. R.~C.,  et~al., 2009, \mn@doi [The Astrophysical Journal]
  {10.1088/0004-637X/703/2/1672}, 703, 1672

\bibitem[\protect\citeauthoryear{King \& Pounds}{King \&
  Pounds}{2015}]{King+15}
King A.,  Pounds K.,  2015, \mn@doi [Annual Review of Astronomy and
  Astrophysics] {10.1146/annurev-astro-082214-122316}, 53, 115

\bibitem[\protect\citeauthoryear{Law et~al.,}{Law et~al.}{2015}]{Law+15}
Law D.~R.,  et~al., 2015, \mn@doi [The Astronomical Journal]
  {10.1088/0004-6256/150/1/19}, 150, 19

\bibitem[\protect\citeauthoryear{Law et~al.,}{Law et~al.}{2016}]{Law+16}
Law D.~R.,  et~al., 2016, \mn@doi [The Astronomical Journal]
  {10.3847/0004-6256/152/4/83}, 152, 83

\bibitem[\protect\citeauthoryear{Lin et~al.,}{Lin et~al.}{2019}]{Lin+19}
Lin L.,  et~al., 2019, \mn@doi [The Astrophysical Journal Letters]
  {10.3847/2041-8213/ab4815}, 884, L33

\bibitem[\protect\citeauthoryear{Lintott et~al.,}{Lintott
  et~al.}{2008}]{Lintott+08}
Lintott C.~J.,  et~al., 2008, \mn@doi [Monthly Notices of the Royal
  Astronomical Society] {10.1111/j.1365-2966.2008.13689.x}, 389, 1179

\bibitem[\protect\citeauthoryear{Lintott et~al.,}{Lintott
  et~al.}{2011}]{Lintott+11}
Lintott C.,  et~al., 2011, \mn@doi [Monthly Notices of the Royal Astronomical
  Society] {10.1111/j.1365-2966.2010.17432.x}, 410, 166

\bibitem[\protect\citeauthoryear{Mallmann et~al.,}{Mallmann
  et~al.}{2018}]{Mallmann+18a}
Mallmann N.~D.,  et~al., 2018, \mn@doi [Monthly Notices of the RAS]
  {10.1093/mnras/sty1364}, \href
  {https://ui.adsabs.harvard.edu/abs/2018MNRAS.478.5491M} {478, 5491}

\bibitem[\protect\citeauthoryear{McAlpine et~al.,}{McAlpine
  et~al.}{2016}]{McAlpine+16}
McAlpine S.,  et~al., 2016, \mn@doi [Astronomy and Computing]
  {10.1016/j.ascom.2016.02.004}, 15, 72

\bibitem[\protect\citeauthoryear{Muzzin et~al.,}{Muzzin
  et~al.}{2013}]{Muzzin+13}
Muzzin A.,  et~al., 2013, \mn@doi [The Astrophysical Journal]
  {10.1088/0004-637X/777/1/18}, 777, 18

\bibitem[\protect\citeauthoryear{Nayakshin \& Zubovas}{Nayakshin \&
  Zubovas}{2012}]{Nayakshin+12}
Nayakshin S.,  Zubovas K.,  2012, \mn@doi [Monthly Notices of the Royal
  Astronomical Society] {10.1111/j.1365-2966.2012.21950.x}, 427, 372

\bibitem[\protect\citeauthoryear{Nelson et~al.,}{Nelson
  et~al.}{2015}]{Nelson+15}
Nelson D.,  et~al., 2015, \mn@doi [Astronomy and Computing]
  {10.1016/j.ascom.2015.09.003}, 13, 12

\bibitem[\protect\citeauthoryear{Nelson et~al.,}{Nelson
  et~al.}{2019}]{Nelson+19}
Nelson D.,  et~al., 2019, \mn@doi [Monthly Notices of the Royal Astronomical
  Society] {10.1093/mnras/stz2306}, 490, 3234

\bibitem[\protect\citeauthoryear{Noeske et~al.,}{Noeske
  et~al.}{2007}]{Noeske+07}
Noeske K.~G.,  et~al., 2007, \mn@doi [The Astrophysical Journal Letters]
  {10.1086/517927}, 660, L47

\bibitem[\protect\citeauthoryear{Osterbrock \& Ferland}{Osterbrock \&
  Ferland}{2006}]{Osterbrock+06}
Osterbrock D.~E.,  Ferland G.~J.,  2006, Astrophysics of gaseous nebulae and
  active galactic nuclei, 2nd. ed. by D.E. Osterbrock and G.J. Ferland.
  Sausalito, CA: University Science Books, 2006

\bibitem[\protect\citeauthoryear{Owen}{Owen}{2007}]{Owen+07}
Owen A.,  2007, \mn@doi [Contemp. Math.] {10.1090/conm/443/08555}, 443

\bibitem[\protect\citeauthoryear{Peterken, Merrifield, {Arag{\'o}n-Salamanca},
  {Fraser-McKelvie}, {Avila-Reese}, Riffel, Knapen  \& Drory}{Peterken
  et~al.}{2020}]{Peterken+20}
Peterken T.,  Merrifield M.,  {Arag{\'o}n-Salamanca} A.,  {Fraser-McKelvie} A.,
   {Avila-Reese} V.,  Riffel R.,  Knapen J.,   Drory N.,  2020, \mn@doi
  [Monthly Notices of the Royal Astronomical Society] {10.1093/mnras/staa1303},
  495, 3387

\bibitem[\protect\citeauthoryear{Rees}{Rees}{1989}]{Rees+89}
Rees M.~J.,  1989, \mn@doi [Monthly Notices of the Royal Astronomical Society]
  {10.1093/mnras/239.1.1P}, 239, 1P

\bibitem[\protect\citeauthoryear{Rembold et~al.,}{Rembold
  et~al.}{2017}]{Rembold+17}
Rembold S.~B.,  et~al., 2017, \mn@doi [Monthly Notices of the Royal
  Astronomical Society] {10.1093/mnras/stx2264}, 472, 4382

\bibitem[\protect\citeauthoryear{Riffel, Pastoriza, {Rodr{\'i}guez-Ardila}  \&
  Maraston}{Riffel et~al.}{2008}]{Riffel+08}
Riffel R.,  Pastoriza M.~G.,  {Rodr{\'i}guez-Ardila} A.,   Maraston C.,  2008,
  \mn@doi [Monthly Notices of the RAS] {10.1111/j.1365-2966.2008.13440.x},
  \href {http://adsabs.harvard.edu/abs/2008MNRAS.388..803R} {388, 803}

\bibitem[\protect\citeauthoryear{Riffel, Pastoriza, {Rodr{\'i}guez-Ardila}  \&
  Bonatto}{Riffel et~al.}{2009}]{Riffel+09}
Riffel R.,  Pastoriza M.~G.,  {Rodr{\'i}guez-Ardila} A.,   Bonatto C.,  2009,
  \mn@doi [Monthly Notices of the RAS] {10.1111/j.1365-2966.2009.15448.x},
  \href {http://adsabs.harvard.edu/abs/2009MNRAS.400..273R} {400, 273}

\bibitem[\protect\citeauthoryear{Riffel, Zakamska  \& Riffel}{Riffel
  et~al.}{2020}]{Riffel+20}
Riffel R.~A.,  Zakamska N.~L.,   Riffel R.,  2020, \mn@doi [Monthly Notices of
  the Royal Astronomical Society] {10.1093/mnras/stz3137}, 491, 1518

\bibitem[\protect\citeauthoryear{Rosario, Burtscher, Davies, Genzel, Lutz  \&
  Tacconi}{Rosario et~al.}{2013}]{Rosario+13}
Rosario D.~J.,  Burtscher L.,  Davies R.,  Genzel R.,  Lutz D.,   Tacconi
  L.~J.,  2013, \mn@doi [The Astrophysical Journal]
  {10.1088/0004-637X/778/2/94}, 778, 94

\bibitem[\protect\citeauthoryear{Rosario, Mendel, Ellison, Lutz  \&
  Trump}{Rosario et~al.}{2016}]{Rosario+16}
Rosario D.~J.,  Mendel J.~T.,  Ellison S.~L.,  Lutz D.,   Trump J.~R.,  2016,
  \mn@doi [Monthly Notices of the Royal Astronomical Society]
  {10.1093/mnras/stw096}, 457, 2703

\bibitem[\protect\citeauthoryear{Rosario et~al.,}{Rosario
  et~al.}{2018}]{Rosario+18}
Rosario D.~J.,  et~al., 2018, \mn@doi [Monthly Notices of the Royal
  Astronomical Society] {10.1093/mnras/stx2670}, 473, 5658

\bibitem[\protect\citeauthoryear{{Ruschel-Dutra}}{{Ruschel-Dutra}}{2020}]{Ruschel-Dutra+20}
{Ruschel-Dutra} 2020, Danielrd6/Ifscube v1.0, Zenodo,
  \mn@doi{10.5281/zenodo.3945237}

\bibitem[\protect\citeauthoryear{Salim, Boquien  \& Lee}{Salim
  et~al.}{2018}]{Salim+18}
Salim S.,  Boquien M.,   Lee J.~C.,  2018, \mn@doi [The Astrophysical Journal]
  {10.3847/1538-4357/aabf3c}, 859, 11

\bibitem[\protect\citeauthoryear{Schaye et~al.,}{Schaye
  et~al.}{2015}]{Schaye+15}
Schaye J.,  et~al., 2015, \mn@doi [Monthly Notices of the Royal Astronomical
  Society] {10.1093/mnras/stu2058}, 446, 521

\bibitem[\protect\citeauthoryear{Smee et~al.,}{Smee et~al.}{2013}]{Smee+13}
Smee S.~A.,  et~al., 2013, \mn@doi [The Astronomical Journal]
  {10.1088/0004-6256/146/2/32}, 146, 32

\bibitem[\protect\citeauthoryear{Speagle, Steinhardt, Capak  \&
  Silverman}{Speagle et~al.}{2014}]{Speagle+14}
Speagle J.~S.,  Steinhardt C.~L.,  Capak P.~L.,   Silverman J.~D.,  2014,
  \mn@doi [The Astrophysical Journal Supplement Series]
  {10.1088/0067-0049/214/2/15}, 214, 15

\bibitem[\protect\citeauthoryear{Springel et~al.,}{Springel
  et~al.}{2005}]{Springel+05}
Springel V.,  et~al., 2005, \mn@doi [Nature] {10.1038/nature03597}, 435, 629

\bibitem[\protect\citeauthoryear{Trussler, Maiolino, Maraston, Peng, Thomas,
  Goddard  \& Lian}{Trussler et~al.}{2020}]{Trussler+20}
Trussler J.,  Maiolino R.,  Maraston C.,  Peng Y.,  Thomas D.,  Goddard D.,
  Lian J.,  2020, \mn@doi [Monthly Notices of the Royal Astronomical Society]
  {10.1093/mnras/stz3286}, 491, 5406

\bibitem[\protect\citeauthoryear{Vazdekis, {S{\'a}nchez-Bl{\'a}zquez},
  {Falc{\'o}n-Barroso}, Cenarro, Beasley, Cardiel, Gorgas  \&
  Peletier}{Vazdekis et~al.}{2010}]{Vazdekis+10}
Vazdekis A.,  {S{\'a}nchez-Bl{\'a}zquez} P.,  {Falc{\'o}n-Barroso} J.,  Cenarro
  A.~J.,  Beasley M.~A.,  Cardiel N.,  Gorgas J.,   Peletier R.~F.,  2010,
  \mn@doi [Monthly Notices of the Royal Astronomical Society]
  {10.1111/j.1365-2966.2010.16407.x}, 404, 1639

\bibitem[\protect\citeauthoryear{Vazdekis, Koleva, Ricciardelli, R{\"o}ck  \&
  {Falc{\'o}n-Barroso}}{Vazdekis et~al.}{2016}]{Vazdekis+16a}
Vazdekis A.,  Koleva M.,  Ricciardelli E.,  R{\"o}ck B.,   {Falc{\'o}n-Barroso}
  J.,  2016, \mn@doi [Monthly Notices of the RAS] {10.1093/mnras/stw2231},
  \href {http://adsabs.harvard.edu/abs/2016MNRAS.463.3409V} {463, 3409}

\bibitem[\protect\citeauthoryear{Vogelsberger et~al.,}{Vogelsberger
  et~al.}{2014}]{Vogelsberger+14}
Vogelsberger M.,  et~al., 2014, \mn@doi [Nature] {10.1038/nature13316}, 509,
  177

\bibitem[\protect\citeauthoryear{Von Der~Linden, Best, Kauffmann  \& White}{Von
  Der~Linden et~al.}{2007}]{VonDerLinden+07}
Von Der~Linden A.,  Best P.~N.,  Kauffmann G.,   White S. D.~M.,  2007, \mn@doi
  [Monthly Notices of the Royal Astronomical Society]
  {10.1111/j.1365-2966.2007.11940.x}, 379, 867

\bibitem[\protect\citeauthoryear{Wake et~al.,}{Wake et~al.}{2017}]{Wake+17}
Wake D.~A.,  et~al., 2017, \mn@doi [The Astronomical Journal]
  {10.3847/1538-3881/aa7ecc}, 154, 86

\bibitem[\protect\citeauthoryear{Walcher, Groves, Budav{\'a}ri  \&
  Dale}{Walcher et~al.}{2011}]{Walcher+11}
Walcher J.,  Groves B.,  Budav{\'a}ri T.,   Dale D.,  2011, \mn@doi
  [Astrophysics and Space Science] {10.1007/s10509-010-0458-z}, 331, 1

\bibitem[\protect\citeauthoryear{Wang \& Loeb}{Wang \& Loeb}{2018}]{Wang+18}
Wang X.,  Loeb A.,  2018, \mn@doi [New Astronomy]
  {10.1016/j.newast.2017.12.004}, 61, 95

\bibitem[\protect\citeauthoryear{Weinberger et~al.,}{Weinberger
  et~al.}{2017}]{Weinberger+17}
Weinberger R.,  et~al., 2017, \mn@doi [Monthly Notices of the Royal
  Astronomical Society] {10.1093/mnras/stw2944}, 465, 3291

\bibitem[\protect\citeauthoryear{Westfall et~al.,}{Westfall
  et~al.}{2019}]{Westfall+19}
Westfall K.~B.,  et~al., 2019, \mn@doi [The Astronomical Journal]
  {10.3847/1538-3881/ab44a2}, 158, 231

\bibitem[\protect\citeauthoryear{Wetzel, Tinker  \& Conroy}{Wetzel
  et~al.}{2012}]{Wetzel+12}
Wetzel A.~R.,  Tinker J.~L.,   Conroy C.,  2012, \mn@doi [Monthly Notices of
  the Royal Astronomical Society] {10.1111/j.1365-2966.2012.21188.x}, 424, 232

\bibitem[\protect\citeauthoryear{White, Bliton, Bhavsar, Bornmann, Burns,
  Ledlow  \& Loken}{White et~al.}{1999}]{White+99}
White R.~A.,  Bliton M.,  Bhavsar S.~P.,  Bornmann P.,  Burns J.~O.,  Ledlow
  M.~J.,   Loken C.,  1999, \mn@doi [The Astronomical Journal]
  {10.1086/301103}, 118, 2014

\bibitem[\protect\citeauthoryear{Yan et~al.,}{Yan et~al.}{2016a}]{Yan+16a}
Yan R.,  et~al., 2016a, \mn@doi [The Astronomical Journal]
  {10.3847/0004-6256/151/1/8}, 151, 8

\bibitem[\protect\citeauthoryear{Yan et~al.,}{Yan et~al.}{2016b}]{Yan+16}
Yan R.,  et~al., 2016b, \mn@doi [The Astronomical Journal]
  {10.3847/0004-6256/152/6/197}, 152, 197

\bibitem[\protect\citeauthoryear{Zhuang \& Ho}{Zhuang \& Ho}{2020}]{Zhuang+20a}
Zhuang M.-Y.,  Ho L.~C.,  2020, \mn@doi [The Astrophysical Journal]
  {10.3847/1538-4357/ab8f2e}, 896, 108

\bibitem[\protect\citeauthoryear{Zhuang, Ho  \& Shangguan}{Zhuang
  et~al.}{2019}]{Zhuang+19}
Zhuang M.-Y.,  Ho L.~C.,   Shangguan J.,  2019, \mn@doi [The Astrophysical
  Journal] {10.3847/1538-4357/ab0650}, 873, 103

\bibitem[\protect\citeauthoryear{Zhuang, Ho  \& Shangguan}{Zhuang
  et~al.}{2020}]{Zhuang+20}
Zhuang M.-Y.,  Ho L.~C.,   Shangguan J.,  2020, arXiv e-prints, 2007,
  arXiv:2007.11285

\bibitem[\protect\citeauthoryear{Zubovas \& Bourne}{Zubovas \&
  Bourne}{2017}]{Zubovas+17a}
Zubovas K.,  Bourne M.~A.,  2017, \mn@doi [Monthly Notices of the RAS]
  {10.1093/mnras/stx787}, \href
  {http://adsabs.harvard.edu/abs/2017arXiv170310782Z} {468, 4956}

\bibitem[\protect\citeauthoryear{Zubovas, Nayakshin, King  \&
  Wilkinson}{Zubovas et~al.}{2013}]{Zubovas+13}
Zubovas K.,  Nayakshin S.,  King A.,   Wilkinson M.,  2013, \mn@doi [Monthly
  Notices of the Royal Astronomical Society] {10.1093/mnras/stt952}, 433, 3079

\bibitem[\protect\citeauthoryear{van~der Wel et~al.,}{van~der Wel
  et~al.}{2014}]{Wel+14}
van~der Wel A.,  et~al., 2014, \mn@doi [The Astrophysical Journal]
  {10.1088/0004-637X/788/1/28}, 788, 28

\makeatother
\end{thebibliography}




 \appendix

\section{Individual maps}

Here we present individual maps for the four high $\Sigma$SFR/M$_\star$ ratio, in left from top to bottom Av$_{gas}$, Av$_{\star}$, Av$_{gas}$/ Av$_{\star}$ and in right side from top to bottom SFR$_{gas}$, SFR$_{\star}$, SFR$_{gas}$/ SFR$_{\star}$ note that a direct comparison of both quantities in individual galaxies is possible because the spaxels have the same area and that the ratio is showing how both quantities compare.

\begin{figure*}
    \centering
     \includegraphics[width=0.85\textwidth,height=0.85\textheight, keepaspectratio]{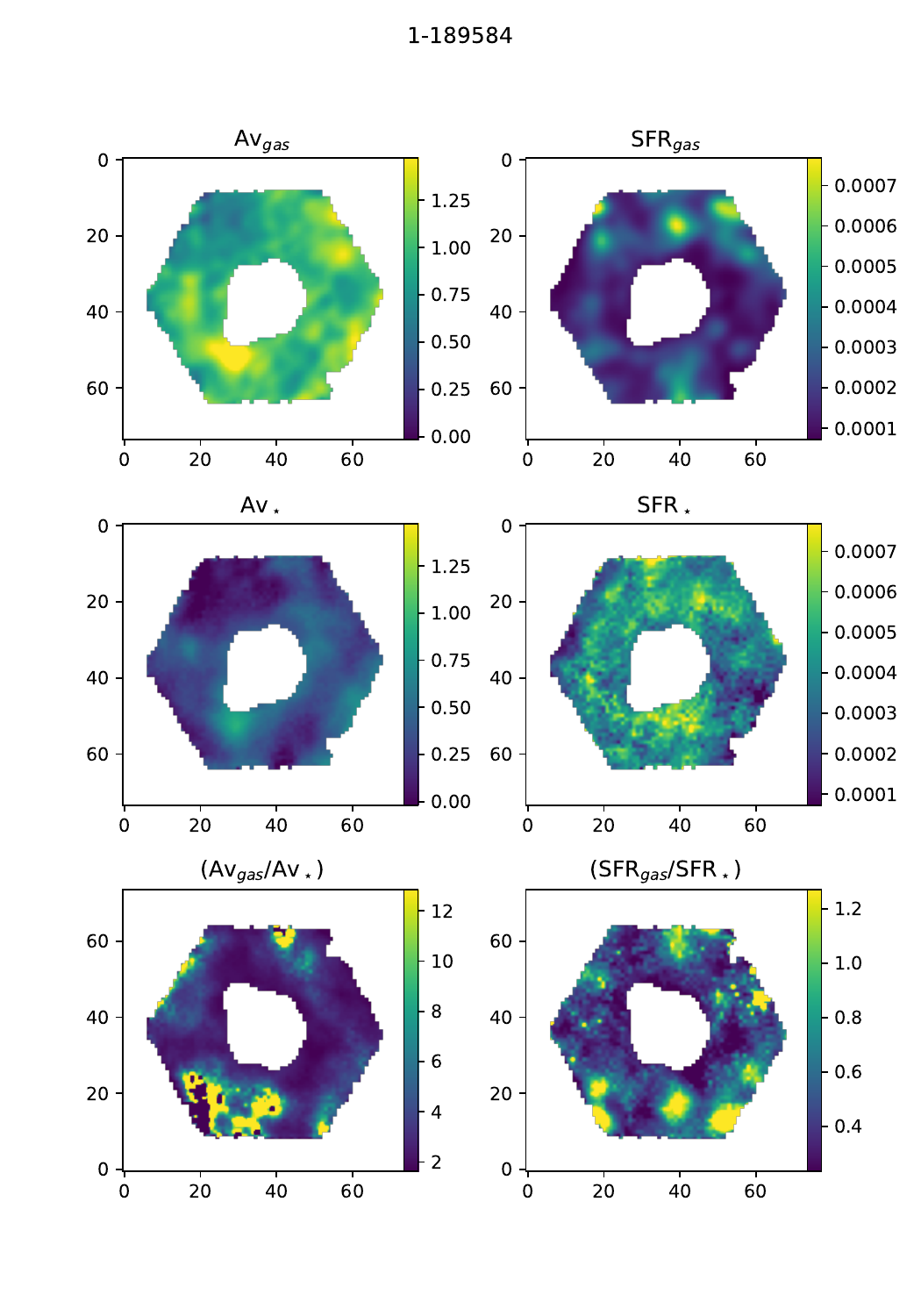}
    \caption{ Individual maps for the four high $\Sigma$SFR/M$_\star$ for MANGAID 1-189584. Left side: from top to bottom A$_{Vg}$, A$_{V\star}$, Av$_{gas}$/ Av$_{\star}$. Right side: from top to bottom SFR$_{g}$, SFR$_{\star}$, SFR$_{g}$/ SFR$_{\star}$. SFR are givens in $M_\odot yr^{-1}$ and A$_V$ in mag.}
    \label{Ap1}
\end{figure*}

\begin{figure*}
    \centering
     \includegraphics[width=0.85\textwidth,height=0.85\textheight, keepaspectratio]{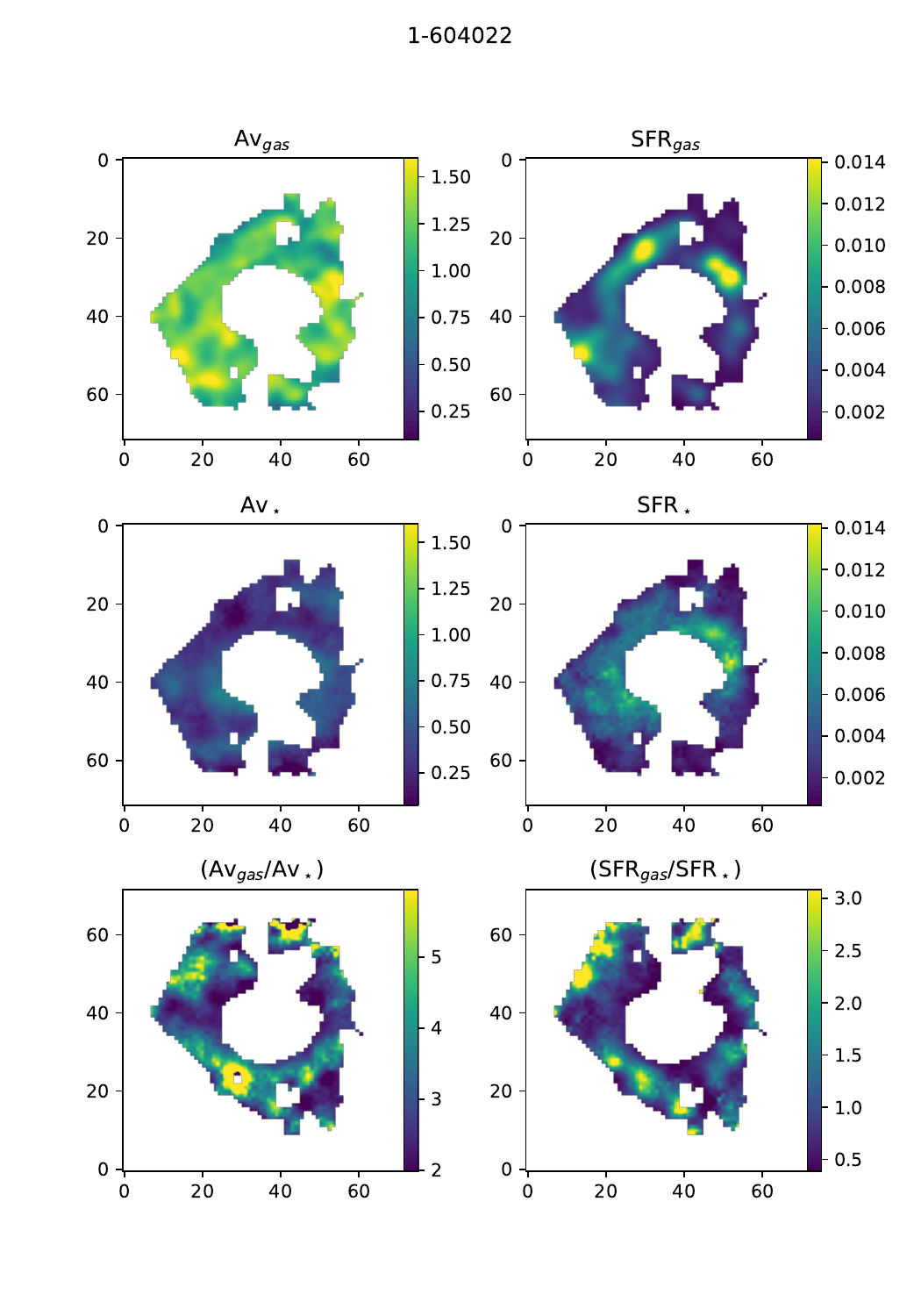}
    \caption{ Same as Fig.~\ref{Ap1} but for 1-604022}
    \label{Ap2}
\end{figure*}

\begin{figure*}
    \centering
     \includegraphics[width=0.85\textwidth,height=0.85\textheight, keepaspectratio]{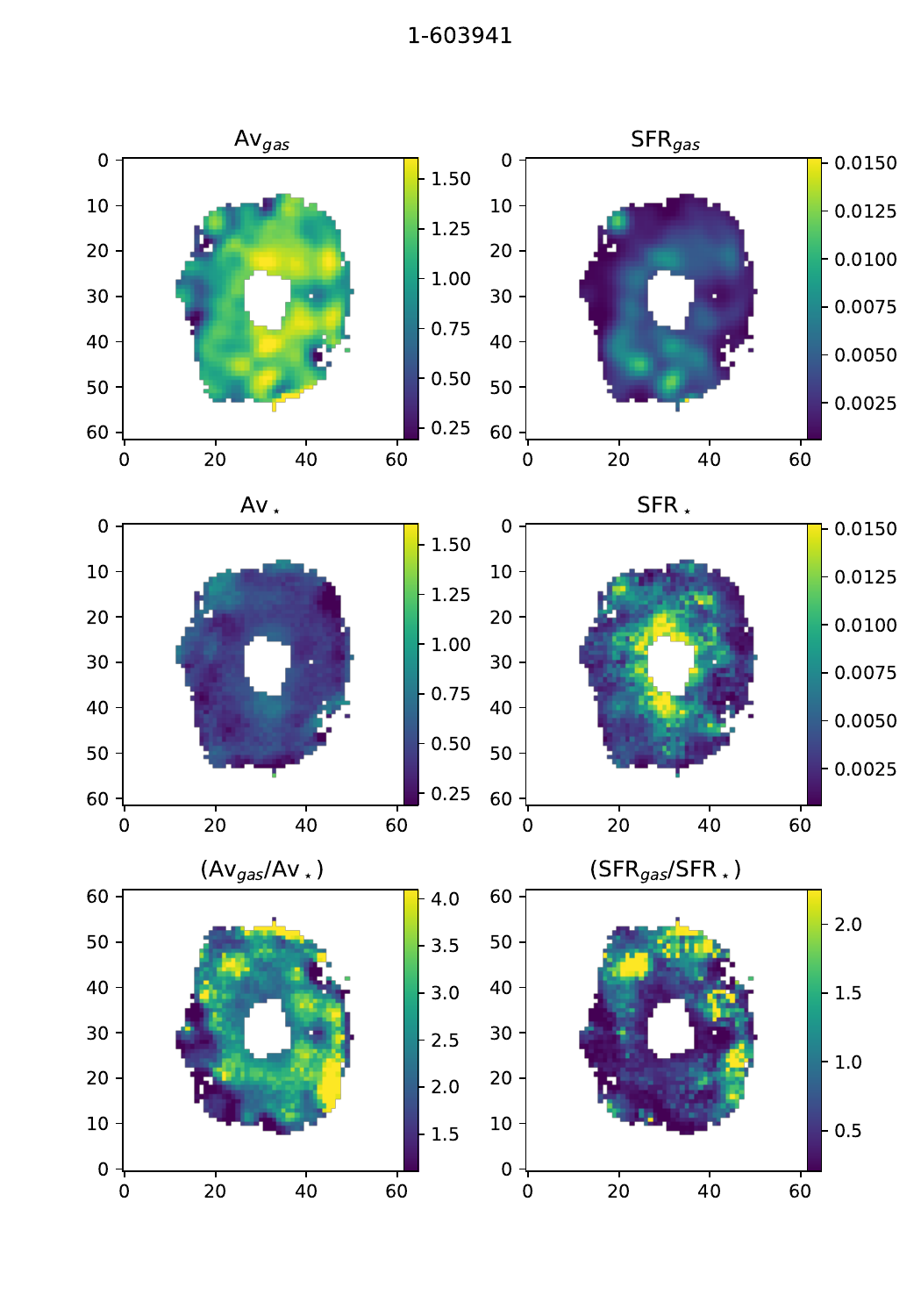}
    \caption{ Same as Fig.~\ref{Ap1} but for 1-603941}
    \label{Ap3}
\end{figure*}

\begin{figure*}
    \centering
     \includegraphics[width=0.85\textwidth,height=0.85\textheight, keepaspectratio]{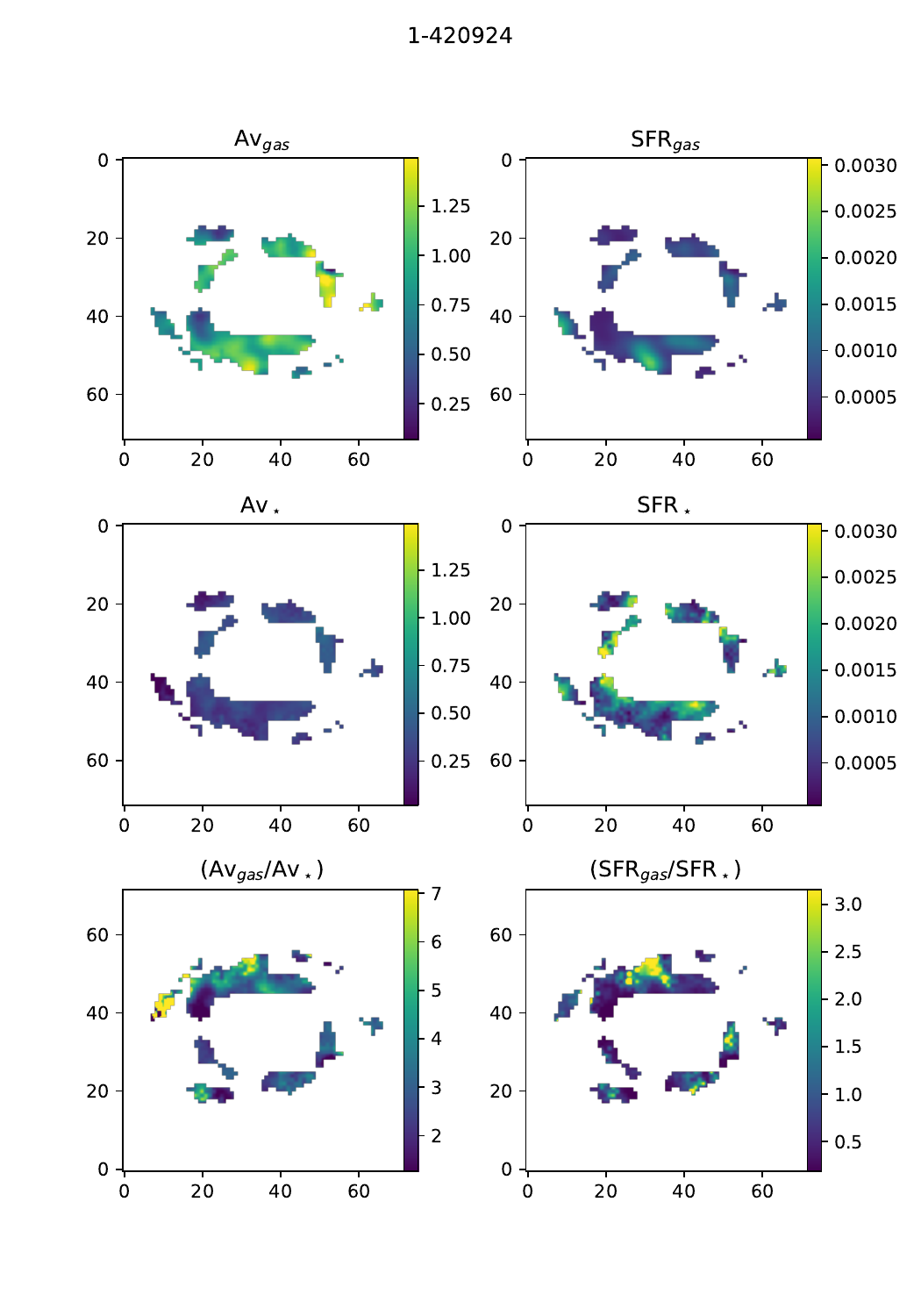}
    \caption{ Same as Fig.~\ref{Ap1} but for 1-420924}
    \label{Ap4}
\end{figure*}



\bsp	
\label{lastpage}
\end{document}